\documentclass[journal]{IEEEtran}
\usepackage{cite}
\usepackage{longtable}
\usepackage{graphicx}
\usepackage{url}
\usepackage[section]{placeins}
\usepackage{float}
\usepackage{makecell}
\usepackage{cuted}
\usepackage[cmex10]{amsmath}
\usepackage{array}
\usepackage{color}
\usepackage{epstopdf}
\usepackage{algorithm}
\usepackage{algorithmic}
\ifCLASSINFOpdf

\DeclareGraphicsExtensions{.pdf,.jpeg,.png}
\else
\fi
\begin{document}
%

\title {Machine Learning in IoT Security: \\Current Solutions and Future Challenges}
\author{
	Fatima Hussain,
	Rasheed Hussain,
    Syed Ali Hassan, and
	Ekram Hossain
	
	\thanks{
		
F. Hussain is with API Operation and Delivery Squad, Royal Bank of Canada, Toronto, Canada (email: fatima.hussain@rbc.com).
		
		R. Hussain is with Networks and Blockchain Laboratory, Innopolis University, Innopolis, Russia (email: r.hussain@innopolis.ru).
		
		S. A. Hassan is with the School of Electrical Engineering and Computer Science (SEECS), National University of Sciences and Technology (NUST), Pakistan (email: ali.hassan@seecs.edu.pk).
		
		E. Hossain is with Department of Electrical and Computer Engineering at University of Manitoba, Winnipeg, Canada (email: Ekram.Hossain@umanitoba.ca).
	}
}
\bibliographystyle{ieeetr}
\maketitle

\begin{abstract}
The future Internet of Things (IoT) will have a deep economical, commercial and social impact on our lives. The participating nodes in IoT networks are usually resource-constrained, which makes them luring targets for cyber attacks. In this regard, extensive efforts have been made to address the security and privacy issues in IoT networks primarily through traditional cryptographic approaches. However, the unique characteristics of IoT nodes render the existing solutions insufficient to encompass the entire security spectrum of the IoT networks. This is, at least in part, because of the resource constraints, heterogeneity, massive real-time data generated by the IoT devices, and the extensively dynamic behavior of the networks. Therefore, Machine Learning (ML) and Deep Learning (DL) techniques, which are able to provide embedded intelligence in the IoT devices and networks, are leveraged to cope with different security problems. In this paper, we systematically review the security requirements, attack vectors, and the current security solutions for the IoT networks. We then shed light on the gaps in these security solutions that call for ML and DL approaches. We also discuss in detail the existing ML and DL solutions for addressing different security problems in IoT networks. At last, based on the detailed investigation of the existing solutions in the literature, we discuss the future research directions for ML- and DL-based IoT security. 

 \end{abstract}

\IEEEpeerreviewmaketitle

\section{Introduction} 

IoT is considered as an interconnected and distributed network of embedded systems communicating through wired or wireless communication technologies \cite{Oscar}. It is also defined as the network of physical objects or \textit{things} empowered with limited computation, storage, and communication capabilities as well as embedded with electronics (such as sensors and actuators), software, and network connectivity that enables these objects to collect, sometime process, and exchange data. The \textit{things} in IoT refer to the objects from our daily life ranging from smart house-hold devices such as smart bulb, smart adapter, smart meter, smart refrigerator, smart oven, AC, temperature sensor, smoke detector, IP camera, to more sophisticated devices such as Radio Frequency IDentification (RFID) devices, heartbeat detectors, accelerometers, sensors in parking lot, and a range of other sensors in automobiles etc. \cite{Fatima}.  There are a plethora of applications and services offered by the IoT ranging from critical infrastructure to agriculture, military, home appliances, and personal health-care \cite{Fuqaha2015}. Furthermore, the domains covered by the IoT services include, but not limited to, energy, building management, medical, retail, transportation, manufacturing, and so on.  The huge scale of IoT networks brings new challenges such as management of these devices, sheer amount of data, storage, communication, computation, and  security and privacy. There have been extensive researches covering these different aspects of IoT (e.g. architecture, communication, protocols, applications, security and privacy) \cite{Fuqaha2015,Granjal2015,Mosenia2017,Lin2017,AMMAR20188,Stellios2018,Restuccia2018,DACOSTA2019,HOU2019,KOUICEM2018,ALABA2017,ZARPELAO2017,MOHAMADNOOR2019,NGUYEN2015,SHA2018}. However, the cornerstone of the commercialization of IoT technology is the security and privacy guarantee as well as consumer satisfaction. The fact that IoT uses enabling technologies such as Software-Defined Networking (SDN), Cloud Computing (CC), and fog computing, also increases the landscape of threats for the attackers. 

Data generated by the IoT devices is massive and therefore, traditional data collection, storage, and processing techniques may not work at this scale. Furthermore, the sheer amount of data can also be used for patterns, behaviors, predictions, and assessment. Additionally, the heterogeneity of the data generated by IoT creates another front for the current data processing mechanisms. Therefore, to harness the value of the IoT-generated data, new mechanisms are needed. In this context, Machine Learning (ML) is considered to be one of the most suitable computational paradigms to provide embedded intelligence in the IoT devices \cite{Mohammad}. ML can help machines and smart devices to infer useful knowledge from the device- or human-generated data. It can also be defined as the ability of a smart device to vary or automate the situation or behavior based on knowledge which is considered as an essential part for an IoT solution. ML techniques have been used in tasks such as classification, regression and density estimation. Variety of applications such as computer vision, fraud detection, bio-informatics, malware detection, authentication, and speech recognition use ML algorithms and techniques. In a similar manner, ML can be leveraged in IoT for providing intelligent services. In this paper, however, we focus on the applications of ML in providing security and privacy services to the IoT networks.         
 
\subsection{Characteristics of IoT Networks} 
In the following we discuss some unique characteristics of IoT networks.

\noindent
\vspace{0.1cm}
\textbf{Heterogeneity}: In an IoT network, a multitude of different devices with different capabilities, characteristics and different communication protocols communicate with each other. More precisely, the devices could use different standards for communication, and different communication paradigms (such as cellular or Ethernet) and variable constraints on the hardware resources. 


\noindent
\vspace{0.1cm}
\textbf{Massive scale deployment}:
It is speculated that the billions of devices connected with each other and through Internet will likely surpass the capabilities of the current Internet. 
The deployment of IoT on massive scale also brings challenges. Some of these challenges include design of networking and storage architecture for smart devices, efficient data communication protocols, proactive identification and protection of IoT from malicious attacks, standardization of technologies, and devices and application interfaces  \cite{Sen,Uday} etc.

\noindent
\vspace{0.1cm}
\textbf{Inter-connectivity}:
 IoT devices are expected to be connected to global information and communication infrastructure and can be accessed from anywhere and anytime. The connectivity depends on the type of service and application provided by the IoT service provider(s). In some cases, the connectivity could be local (such as in case of connected car technology or swarm of sensors) whereas in other cases it could be global such as in case of smart home access through mobile infrastructure and critical infrastructure management. 
   
  \noindent
  \vspace{0.1cm}
\textbf{Communication in close proximity}:
  Another salient feature of IoT is the communication in close proximity without involving the central authorities such as base stations. Device-to-Device communication (D2D) leverages the characteristics of point-to-point communication such as Dedicated Short Range Communication (DSRC) and similar technologies. The architecture of traditional Internet is more inclined towards network-centric communication whereas recently the decoupling of networks and services also enable device-centric as well as content-centric communication which enriches the IoT service spectrum.

\noindent
 \vspace{0.1cm}
\textbf{Ultra-Reliable and Low Latency Communication (URLLC)}:
 This property of IoT networks is required in critical real-time applications such as industrial process automation, remote surgery, and intelligent traffic transport system, where the major performance constraints are both delay and reliability. 
  
\noindent
\vspace{0.1cm}
\textbf{Low-power and low-cost communication}:
Massive connectivity of IoT devices requires ultra low-power and low cost solutions for efficient network operations. 

\noindent
\vspace{0.1cm}
\textbf{Self-organization and self-healing characteristics}:
These are required for urgent and contemporary IoT communication that includes emergency or disaster situations. In such situations, reliance on the network infrastructure is not an option and therefore, self-organizing networks should be deployed. 
 
\noindent
\vspace{0.1cm}
\textbf{Dynamic changes in the network}:
IoT consists of massive number of devices that need to be managed in an efficient way. These devices will act dynamically, for instance the sleep/wakeup time of devices will depend on the application, when do these device use Internet and when do they communicate directly, and so on. These characteristics must also be incorporated into the IoT networks. 

\noindent
\vspace{0.1cm}
\textbf{Safety}:
Among other characteristics, safety is also of paramount importance for the smooth functionality of the IoT networks. Safety is considered for both the consumers and devices because the large number of IoT devices connected to Internet may jeopardize the personal data that is shared through these devices. Furthermore, privacy and the security of device itself is also an important factor. 

\noindent
\vspace{0.1cm}
\textbf{Intelligence}:
One of the most intriguing characteristics of IoT is the intelligence through which timely and informed decisions are made. The data generated by IoT devices should be processed in a way to make real sense out of it and perform actions as a results of decisions made on the basis of the processed data. 

\subsection{Security Challenges in IoT Deployment}

Security and privacy are two of the main factors in the commercial realization of the IoT services and applications. Current Internet is the luring playground for security attacks ranging from simple hacks all the way to corporate level well-coordinated security breaches that have adversely affected different industries such as health-care and business. The limitations of the IoT devices and the environment they operate in, pose additional challenges for the security of both applications and the devices. To date, security and privacy issues have been extensively researched in the IoT domain from different perspectives such as communication security, data security, privacy, architectural security, identity management, malware analysis, and so on \cite{Granjal2015}. Detailed discussion regarding security challenges and threat model follows in Section \ref{sec:securityissues}.

\subsection{Gaps in the Existing Security Solution for IoT Networks}
For the successful realization IoT, it is important to analyze the roots of the security and privacy issues. More precisely the term IoT has been tossed from the existing technologies and therefore, it is imperative to know whether the security challenges in IoT are new or a rehash of the inheritance from the old technologies. Fernandes et al. \cite{Fernandes2017} focused on similarities and differences of the security issues in IoT and the traditional IT devices. Furthermore, they also focused on the privacy issues. The main driving factors to argue on the similarities and differences include software, hardware, network, and applications. Based on these classifications, there are fundamental similarities between the security issues in traditional IT domain and the IoT. However, the primary concern of the IoT is the resource-constraints that hinder the adaptation of already available sophisticated security solutions in IoT networks. Furthermore, solutions to the security and privacy issues in IoT require cross-layer design and optimized algorithms. For instance due to computational constraints, IoT devices may need new breed of optimized cryptographic and other algorithms to cope with security and privacy. On the other hand, the number of devices in IoT pose other challenges for the security mechanisms. 

Most of the security challenges are complex and the solutions cannot be discrete. For instance, in case of security challenges such as DDoS or intrusion, there is a probability of false positives which will render the solutions to be ineffective against these attacks. Additionally, it will also decrease the consumer trust and thereby degrading the effectiveness of these solutions. Therefore, a holistic security and privacy approach towards IoT will have nominations from the existing security solutions as well as development of new intelligent, robust, evolutionary, and scalable mechanisms to address security challenges in IoT. 

\subsection{Machine Learning: A Solution to IoT Security Challenges}

Machine learning refers to intelligent methods  used to optimize performance criteria using example data or past experience(s) through learning. More precisely, ML algorithms build models of behaviors using mathematical techniques on huge data sets. ML also enables the ability (for smart devices) to learn without being explicitly programmed. These models are used as a basis for making future predictions based on the newly input data. ML is interdisciplinary in nature and inherits its roots from many disciplines of science and engineering that include artificial intelligence, optimization theory, information theory, and cognitive science, to name the few \cite{Junfei}. Machine learning is utilized when human expertise either do not exist or cannot be used such as navigating a hostile place where humans are unable to use their expertise, for instance robotics, speech recognition etc. It is also applied in situations where solution to some specific problem changes in time (routing in a computer network or finding malicious code in a software or application). Furthermore, it is used in practical smart systems, for instance Google uses ML to analyze threats against mobile endpoints and applications running on Android. It is also used for identifying and removing malware from infected handsets. Likewise, Amazon has launched a service {\bf Macie} that uses ML to sort and classify data stored in its cloud storage service. Although ML techniques perform well in many areas; however, there is a chance of false positives and true negatives. Therefore, ML techniques need guidance and modification to the model if inaccurate prediction is made. On the contrary, in Deep Learning (DL), a new breed of ML, the model can determine the accuracy of prediction by itself.  Due to self-service nature of DL models, it is rendered as more suitable for classification and prediction tasks in innovative IoT applications with contextual and personalized assistance.

Although traditional approaches are widely used for different aspects of IoT (e.g. applications, services, architectures, protocols, data aggregation, resource allocation, clustering, analytics) including security, the massive scale deployment of IoT however, advocates for intelligent, robust, and reliable techniques. To this end, ML and DL are promising techniques  for IoT networks due to several reasons, e.g. IoT networks produce sheer amount of data which is required by ML and DL approaches to bring intelligence to the systems. Furthermore, the utility of the data generated by the IoT is better utilized with the ML and DL techniques which enable the IoT systems to make informed and intelligent decisions. ML and DL are largely used for security, privacy, attack detection, and malware analysis. DL techniques can also be used in IoT devices to perform complex sensing and recognition tasks to enable the realization of new applications and services considering real-time interactions among humans, smart devices and physical surroundings. Some of the security related real-world applications of ML are as follows:

\begin{itemize}
  \item Face recognition for forensics: pose, lighting, occlusion (glasses, beard), make-up, hair style, etc.
  \item Character recognition for security encryption: different handwriting styles.
  \item Malicious code identification: identifying malicious code in applications and software. 
  \item Distributed Denial of Service (DDoS) detection: detecting DDoS attacks on infrastructure through behavior analysis.
\end{itemize}

Using ML and DL techniques in IoT applications on the other hand bring along new challenges. These challenges are multi-faceted. For instance, it is challenging to develop a suitable model to process data from diverse IoT applications. Similarly, labeling input data effectively is also a cumbersome task. Another challenge is using minimum labelled data in the learning process. Other challenges stems from the deployment of these models on resource-constrained IoT devices where it is essential to reduce the processing and storage overhead \cite{Shuochao}. Similarly, critical infrastructure and real-time applications cannot withstand the anomalies created because of ML or DL algorithms. In the above context, it is imperative to systematically review the security solutions of IoT that leverage ML and DL. 

\subsection{Existing Surveys}
IoT has a rich literature and to date, many surveys have been published that cover different aspects of the IoT security. In this section, we summarize the existing surveys and compare them with our work. To the best of our knowledge, most of the surveys we found in the literature do not focus on the ML techniques used in IoT. Furthermore, the existing surveys are either application-specific or do not encompass the full spectrum of the security and privacy in the IoT networks. Table~\ref{table:surveys} summarizes the existing surveys in the literature that cover different aspects of the IoT. In this table, we outline the topics covered in these surveys and the respective enhancements in our survey. 

The current literature covers the security in IoT by investigating the existing traditional solutions and the solutions provided through the new emerging technologies. However,  surveys covering ML- and DL-based solutions do not exist. Although ML and DL have been covered in few surveys but the overall information on the comprehensive usage of ML and DL is scarce. To fill the gaps, we conduct a comprehensive survey of the ML and DL techniques used in IoT security. 

\subsection{Scope of This Survey and Contributions}

\begin{figure*}
\centering
\includegraphics[width=1\textwidth]{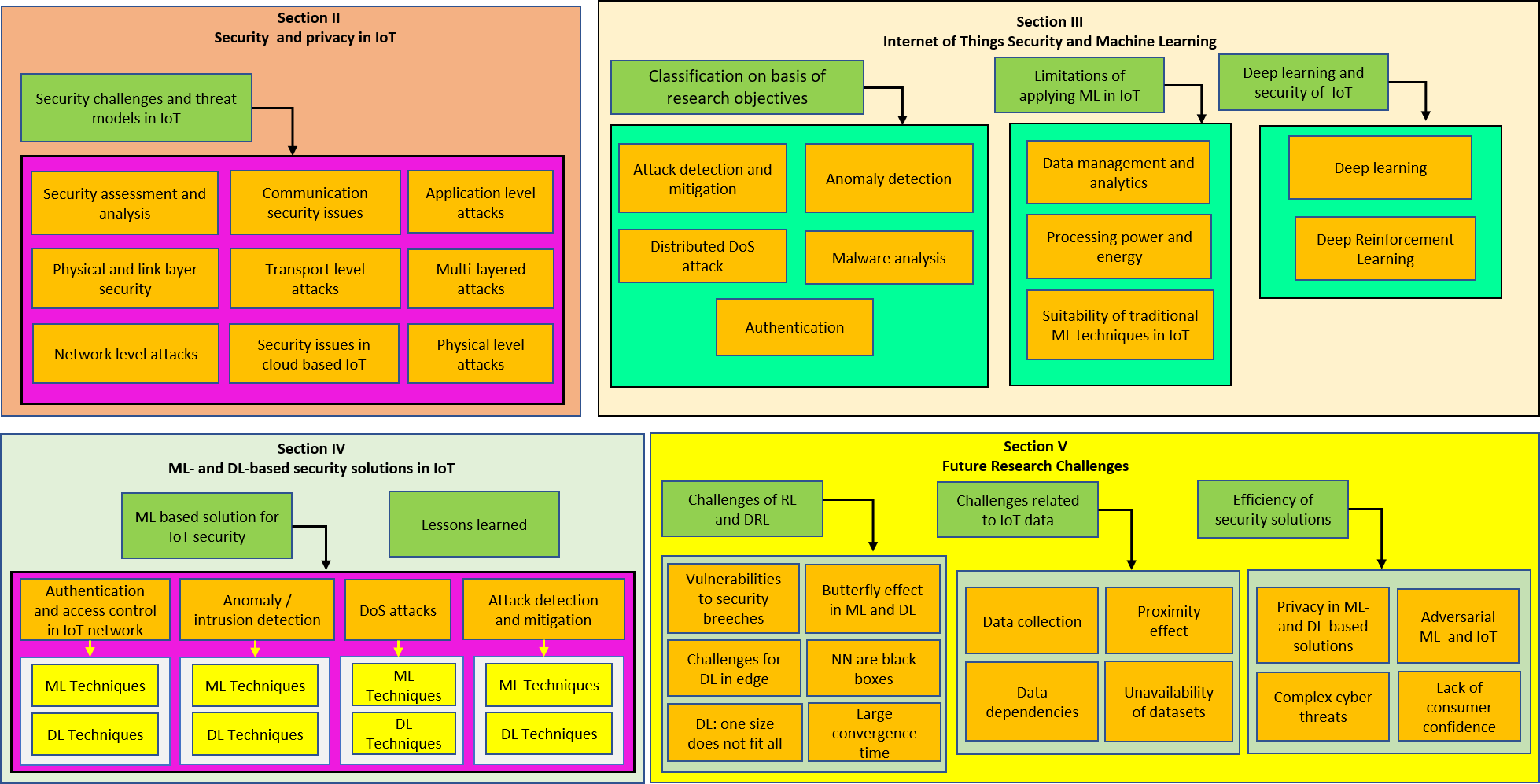}
\caption{Taxonomy of the survey.}
\label{fig:taxonomy}
\end{figure*}

We carry out in-depth systematic survey of ML- and DL-based security solutions in IoT.  First we discuss  the security requirements of IoT applications, threats and attacks in IoT networks. 
Then we discuss the role of ML and DL in IoT and discuss different ML and DL techniques that are actively leveraged for IoT applications and services. To focus more on the functional side of the IoT, we dive deeper into the ML- and DL-based security solutions in IoT.  To this end, we also discuss the existing research challenges and future directions for more research on the ML and DL for IoT networks. Our aim is to bridge the gap between the requirements of the IoT security and the capabilities of the ML and DL which will help addressing the current security challenges of the IoT networks. We pictorially illustrate the scope of this survey in Fig. \ref{fig:taxonomy}. 

The main contributions of this paper can be summarized as follows: 
\begin{enumerate}
\item We present an in-depth systematic and comprehensive survey of the role of machine and deep learning mechanisms in IoT.
\item We describe state-of-the-art results on ML and DL in IoT network with a focus on security and privacy of the IoT networks.
\item We describe the limitations of the existing security solutions of the IoT networks that lead to using ML and DL techniques.  
\item We also present an in-depth review of different research challenges related to the application of ML and DL techniques in IoT that need to be addressed.

\end{enumerate}

{\em To the best of our knowledge and based on our extensive search in the literature, this is the first such effort to review IoT from ML and DL standpoint. We carry out an in-depth investigation of the security and privacy in the IoT networks. In essence, we investigate in detail, the security requirements, attack surface in IoT, and then discuss the ML and DL-based solutions to mitigate the security attacks in IoT networks.} {\em It is worth mentioning that we have covered the existing surveys till 2019. Furthermore, our survey contains the recent works carried out in the fields of ML and DL for IoT networks.}

The rest of the paper is organized as follows: Table \ref{table:acronyms} lists all the acronyms used in the paper. We discuss threat model of IoT networks in Section \ref{sec:securityissues}. In Section \ref{sec:MLDL}, we discuss the role of ML in the IoT networks and briefly review different ML and DL techniques.  We survey the existing ML- and DL-based solutions for IoT networks in Section \ref{sec:MLSecSolutions}. The future research directions are discussed in Section \ref{sec:challenges} and Section \ref{sec:conclusion} concludes this paper. 

\section{Security Challenges and Threat Models in IoT}
\label{sec:securityissues}

In essence, IoT employs a transformative approach to provide consumers with numerous applications and services. The pervasive deployment of large number of devices also increase the attack surface. On the other hand, the fact that IoT devices are (usually) resource-constrained; therefore, it is not feasible to use sophisticated security mechanisms against notorious attacks. Furthermore, it is worth mentioning that the original Internet was not designed for IoT. Therefore, it is imperative to provide IoT security on top of the existing security mechanisms of the Internet and the underlying technologies. To this end, IoT uses different communication technologies such as, but not limited to, IPv6, Zigbee, 6LoWPAN, Bluetooth, Z-Wave, WiFi, and Near Field Communications (NFC), to name a few \cite{Sarawi2017}. These aforementioned communications technologies have their own shortcomings and limitations from security standpoint and these limitations are inherited in the IoT domain as well. In addition to these issues, the underlying TCP/IP based communication infrastructure is prone to  challenges such as scalability, complexity, addressing, configuration, and insufficient resource utilization that limit the possibilities of using it for diverse and heterogeneous networks such as IoT \cite{Liu2018}. To this end, different alternative technologies such as information-centric networking (ICN) and software-defined networking (SDN) have been utilized to serve as underlying communication infrastructures for IoT \cite{Chen2016,Liu2018}. Here, we provide the summary of the threats and attacks faced by IoT. Without loss of generality, security attacks in IoT can be abstractly divided into physical, network, transport, application, and encryption attacks.  

 \subsection{Physical Attacks} 
In physical attacks, the attackers have direct access to the devices and manipulate different aspects of the devices. To get access to the physical devices, social engineering is one of the most prominent methods where the attackers access the devices and perform real attack that ranges from physical damage to the device to eavesdropping, side-channels, and other related attacks \cite{Andrea2015,Benzarti2017}. Despite the fact that different technologies are used at the physical layer for IoT, the nature of physical attacks mostly resemble and need social engineering-like approaches. Furthermore, to launch physical attacks, the attackers must be in the close proximity of the devices/hardware with different intentions such as physically destroying the hardware, limiting its lifetime, endangering the communication mechanism, tampering with the energy source, and so on. It is also worth noting that physical attacks maybe stepping stone for other attacks, for instance disabling an alarm in a home could lead to a burglary or other related damage in smart home environment. Similarly, a replacement of sensor with a malicious sensor would lead to sensitive data leakage. Injection of malicious node into the network can also cause man-in-the-middle attack that enables that attacker to escalate privileges and launch other attacks. Furthermore, such tampering with devices may also enable the attackers to make changes in the routing tables and security keys that will affect the communication with upper layers \cite{Perrig2004,Li2013,FINOGEEV20176}. Other physical attacks include jamming radio frequencies which denies the communication in IoT environment. Among many other repercussions, jamming causes denial of service in IoT thereby adversely affecting the functionality of IoT applications \cite{Salameh2018,Hu2018}. As has been mentioned, the attackers also use different social engineering approaches to have physical access to hardware/devices for different purposes such as the attacks we already mentioned. Through social engineering, the attackers may manipulate users to gain physical access to the devices \cite{Sadeghi2015,Patil2014}. 

\subsection{Physical (PHY) and Link Layer Security Issues}
IoT combines various communication technologies at the lower layers of TCP/IP protocol stack and thus-forth provides a complex heterogeneous network. These technologies include, but not limited to, ZigBee, WSN, MANET, WiFi, RFID, NFC, and so forth and furthermore these technologies have their own security issues. In this subsection, we will put light on the security issues in the physical and data link layers of IoT.  Yang et al. \cite{Yang2017} and Granjal et al. \cite{Granjal2015} discussed in detail, the security issues in IoT at different layers and their existing solutions. As has been mentioned before, the heterogeneity is introduced at physical layer of the IoT and then different amendments are made at data link layer, for instance special channel design and so forth, depending on the underlying physical layer technology. To this end, the security mechanisms of IoT must encompass the heterogeneity at the physical and data link layer. 

There are different security issues in physical layer of IoT depending on the underlying technology, for instance in case of sensor nodes, physical attacks on sensor nodes must be mitigated. Furthermore, detection of malfunction in the hardware is also of paramount importance and must be handled to avoid anomalies at upper layer. Intrusion is another physical security issue that needs efficient countermeasures from both detection and prevention standpoint. 
It is worth noting that there could be many attack vectors even for intrusion detection at upper layers, for instance routing attack \cite{RAZA2013}. From fault detection standpoint, it is important at par to detect the faulty nodes in IoT because it directly affects the quality of service (QoS) of the IoT application \cite{Jadav2017,Nishiguchi2018}. 

The main objective of IoT is to provide a common ground for low-energy constrained devices (aka things) to co-exist at the lower layers (PHY and Link layers) and provide a common communication platform for heterogeneous devices. To this end, IEEE has mandated a standard known as IEEE 802.15.4 to allow constrained device communicate efficiently \cite{Khanafer2014}. The fact that in IoT, higher layers use low power protocols such as 6LoWPAN and  Constrained Application Protocol (CoAP), a mechanism is needed at lower layers to enable these protocols work seamlessly. In this context, IEEE 802.15.4 provides the necessary amendments at the lower layers for these protocols. It is worth noting that, IEEE 802.15.4 incorporates security only at the data link layer. Data link or MAC layer is responsible for the channel access for different devices in addition to frame validation, access management, time management, and security. Here we particularly focus on the security at this layer provided by IEEE 802.15.4.  The security provided by this standard at MAC layer does not only make sure that the node level data transmission is secure, but also complements the security of upper layers. 
In this regard, symmetric cryptography algorithms such as AES are proved to be fast and efficient when implemented on chip, therefore such implementation in IEEE 802.15.4 hardware will compliment the lower layer security \cite{Yicheng2008, Bansod2015}. IEEE 802.15.4 standard implements AES algorithm and different implementations have been proposed to deal with the resource constraints of the devices \cite{Oukili2017}. The standard supports different security modes at the link layer, for instance the data may not be encrypted with only integrity check, or the data may be encrypted along with the integrity check. 

\begin{table*}
\caption{Existing Surveys}
\label{table:surveys}
\begin{tabular}{|m{0.8cm}|m{0.8cm}|m{5.5cm}|m{2.3cm}|m{5.8cm}|}

\hline 
\textbf{Year} & \textbf{Paper} & \textbf{Topic(s) of the survey}  & \textbf{Related sections in our paper}  & \textbf{Enhancements in our paper}  \\ 
\hline
2015  & \cite{Granjal2015}  & Communication security protocols & Sec. \ref{sec:securityissues} & Security requirements, threats, vulnerabilities, and ML- and DL-based solutions  \\ 
\hline
2017 & \cite{Yang2017} & IoT authentication and access control & Sec. \ref{sec:securityissues} & Detailed security and privacy solutions through ML and DL in IoT \\
\hline
2017 & \cite{Mosenia2017} & Security in the edge layer of IoT & Sec. \ref{sec:securityissues} & Coverage of entire IoT from security, privacy standpoint \\
\hline
2017 & \cite{Lin2017} & System architecture, and security, privacy in edge-/fog-based IoT & Sec. \ref{sec:securityissues} & Enhanced coverage of state-of-the-art ML- and DL-based security and privacy in generic IoT \\
\hline
2018 & \cite{AMMAR20188} & Security of IoT framework architectures & Sec. \ref{sec:securityissues} & In-depth coverage of security and privacy in generic IoT with focus on state-of-the-art ML and DL techniques \\
\hline
2018 & \cite{Stellios2018} & IoT-enabled attacks on different sectors and assess different attacks in critical infrastructure & Sec. \ref{sec:securityissues} & In-depth coverage of security issues, threats, attacks, and solutions in generic IoT \\
\hline
2018 & \cite{Restuccia2018} & Security threats in IoT & Sec. \ref{sec:securityissues} & Enhanced threats, attacks, and solutions in IoT \\
\hline
2019 & \cite{DACOSTA2019} & ML-based techniques for IDS in IoT & Sec. \ref{sec:MLSecSolutions} & Coverage of security and privacy and ML-based techniques in IoT \\
\hline
2019 & \cite{HOU2019} & Data security in IoT and data lifecycle & N/A & Covering in-depth security issues and their ML- and DL-based solutions in IoT \\
\hline
2017 & \cite{ALABA2017} & Threats and vulnerabilities in IoT applications, architecture and possible attacks & Sec. \ref{sec:securityissues} & Enhanced threat landscape, requirements, attacks, and their respective solutions in generic IoT security spectrum \\
\hline
2017 & \cite{ZARPELAO2017} & IDS in IoT, detection methods, placement and validation strategies & Sec. \ref{sec:securityissues} & Detailed coverage of security and privacy issues and state-of-the-art based on ML and DL \\
\hline
2019 & \cite{ASGHARI2019} & Security of IoT applications in different domains & Sec. \ref{sec:securityissues} & Coverage of generic IoT applications with solutions, independent of particular domains \\
\hline
2019 & \cite{MOHAMADNOOR2019} & Current development in IoT security, challenges, simulators, and tools & Sec. \ref{sec:securityissues} and \ref{sec:MLSecSolutions} & In-depth and more detailed survey of the security requirements, threats, attacks, and solutions in IoT \\
\hline
2016 & \cite{AIREHROUR2016} & Secure routing protocols in IoT & N/A & Focus on the applications security and privacy \\
\hline
2015 & \cite{NGUYEN2015} & Security protocols and key distribution & Sec. \ref{sec:securityissues} & Focus on security in a holistic way \\
\hline
2018 & \cite{SHA2018} & Security challenges in IoT and sensor networks & Sec. \ref{sec:securityissues} & Detailed security challenges, attacks, and solutions in IoT \\
\hline
2017 & \cite{GUO2017} & Trust models for service management in IoT & N/A & Coverage of different aspects of security with solutions based on ML and DL \\
\hline 
2018 & \cite{COLAKOVIC2018} & Open issues and challenges in IoT and enabling technologies & Sec. \ref{sec:challenges} & Enhanced state-of-the-art and research challenges in ML-driven IoT security \\
\hline
2018 & \cite{Mohammadi2018} & IoT data analytics through DL & N/A & In-depth review of ML- and DL-based security solutions in IoT\\
 \hline
 \end{tabular}
\end{table*}

\begin{table*}
\caption{Acronyms and their explanations}
\label{table:acronyms}
\begin{tabular}{|m{2.5cm}|m{5.5cm}|m{2.5cm}|m{5.5cm}|}
\hline 
\textbf{Acronym} & \textbf{Explanation}  & \textbf{Acronym}  & \textbf{Explanation}  \\ 
\hline
\hline
ML & Machine Learning & RFID & Radio Frequency IDentification \\
 \hline
 UURLLC & Ultra-Reliable and Low Latency Communication & DL & Deep Learning \\
 \hline
 RL & Reinforce Learning & SVM & Support Vector Machine \\
 \hline
 SVR & Support Vector Regression & KNN & K-Nearest Neighbour \\
 \hline
 NB & Naive Bayes & NN & Neural Network \\
 \hline
 DNN & Deep Neural Network & CNN & Convolutional Neural Network \\
 \hline
 PCA & Principal Component Analysis & RNN & Recurrent Neural Network \\
 \hline
 MLP & Multi-Layer Perception & ELM & Extreme Learning Machine \\
 \hline
 ESFCM & ELM-based Semisupervised Fuzzy C-Means & ANN & Artificial Neural Network \\
 \hline
 LSTM & Long-Short Term Memory & DRL & Deep Reinforcement Learning \\
 \hline
 QoS & Quality of Service & CSI & Channel State Information \\
 \hline
  SDN & Software-Defined Network & D2D & Device-to-Device Communication \\
 \hline
 NFC & Near-Field Communication & ICN & Information-Centric Networking \\
 \hline
 DDoS & Distributed Denial of Service & OWASP & Open Web Application Security Project \\
\hline
RBAC & Role-Based Access Control & CWAC & Context-Aware Access Control \\
\hline
PBAC & Policy-Based Access Control & ABAC & Attribute-Based Access Control \\
\hline
UCAC & Usage Control-based Access Control & CAC & Capability-based Access Control \\
\hline
OAC & Organizational-based Access Control & XACML & eXtensible Access Control Markup Language \\
\hline
OAuth & Open Authentication & UMA & User-Managed Access \\
\hline
CoAP & Constrained Application Protocol & FCM & Fuzzy C-Means \\
\hline
MCA & Multivariate Correlation Analysis & SINR & Signal-to-Interference Noise Ratio \\
\hline
CPS & Cyber Physical System & IRG & Influential Relative Grade \\
\hline
LDA & Linear Discriminant Analysis & RaNN & Random Neural Networks \\
\hline
IMA & Illegal Memory Access & XSS & Cross-Site Scripting \\
\hline 
DEL & Deep Eigenspace Learning & WMS & Wireless Multimedia System \\
\hline
BAN & Body Area Network & AML & Adversarial Machine Learning \\
 \hline
 \end{tabular}
\end{table*}

\subsection{Network Layer Attacks}
At the network level, the attacks are aimed at routing, data and traffic analysis, spoofing, and launching man-in-the-middle attack. Besides, sybil attacks are also possible at the network layer where fake identities/sybil identities are used to create illusions in the network \cite{Hussain2014,Dong2015}. 
On the other hand, intrusion through different means, provides a way-in for the attacker to the system where the attackers can launch plethora of other attacks and therefore, securing the network is essential to contain the attacks at early stages. At the network layer, the attacker can also leverage a compromised node to use it as fake forwarding node and create a sinkohole. This type of attack, usually associated with sensor networks and mobile ad hoc networks, is equally dangerous in IoT environment~\cite{Han2015}. With these attacks, the possibility of launching a collaborative DDoS attacks increases, and thereby disrupting the whole IoT network. At the network layer, the attacker can achieve this by bombarding the network with more traffic through compromised nodes than the network can handle.  Compromising IoT nodes and masquerading identities will have catastrophic consequences on the network because with such fake nodes (either non-existent fake identities or existing compromised identities) enable the attackers to launch sybil attacks where sybil nodes (fake nodes) give the illusion to the core network as if real nodes were sending data. 
To summarize, the attack vectors in network layer target the communication aspects of the IoT and exploit the resource constraints and lack of sophisticated authentication and authorization schemes.

\subsection{Transport Layer Attacks}
Transport layer is responsible for process to process delivery where transport protocols enable the processes to exchange data. In the context of IoT, the traditional transport layer security issues still persist. The most serious attack at this layer is the denial of service attack that chokes the network and results in denial of services to the applications. It is worth mentioning that due to the nature of IoT, traditional TCP and UDP protocols do not scale with resource-constrained devices, and therefore lightweight versions of transport protocols have been proposed in the literature \cite{Katsikeas2017,Mondal2017,Haroon2017}. However, the security of these protocols is of primary importance to alleviate the DoS and DDoS attacks in IoT. 

\subsection{Application Layer Attacks}
IoT applications are relatively lucrative targets for the attackers because applications level attacks are relatively easy to launch. Some of the well-known attacks include, but not limited to, buffer overflow attacks, malware attacks, denial of services, phishing, exploiting the WebApp vulnerabilities, cryptographic attacks, side channel attacks, and man-in-the-middle attacks. Buffer overflows are one of the mostly used attack vectors in different applications \cite{Buddhika2016}. Existing techniques to mitigate buffer overflow mechanism include static and dynamic code analysis, and other sophisticated mechanisms such as symbolic debugging; however, these techniques cannot be used with IoT due to resource constraints. IoT applications are also prone to malicious code injection as a result of buffer overflow and other vulnerabilities such as SQL injection, cross-site scripting, object referencing, and so forth. Open Web Application Security Project (OWASP) identified the top 10 vulnerabilities that causes different attacks on applications. The latest list of most widely found vulnerabilities was compiled by OWASP in 2017\footnote{\url{https://www.owasp.org/index.php/Top_10-2017_Top_10}}. These vulnerabilities lead to range of other attacks that could be launched, for instance, injection of malicious code, phishing, access control, privilege escalation, and so on. Furthermore, through malicious code injection attacks, the attacker can steal sensitive information, tampering with information, and range of other malicious activities \cite{Heer2011}. Furthermore, botnets are other serious threats to IoT infrastructure and the applications \cite{Bertino2017}. Mitigation of attacks launched through intelligent botnets is a serious challenge for IoT because such botnets intelligently scan and crawl through the network looking for known vulnerabilities and exploiting them to launch different attacks such as massive DDoS. It is also worth mentioning that due to resource constraints, sophisticated cryptographic protocols are not feasible for IoT devices which leaves them at the mercy of capable attackers to launch cryptographic attacks. 
In a nutshell, the attack surface at application layer of the IoT infrastructure is wide and relatively computationally expensive to mitigate.   

\subsection{Multi-Layered Attacks}
In addition to the aforementioned attacks, there are multi-layered attacks that could be launched on IoT infrastructure. These attacks include  traffic analysis, side-channel attacks, replay attacks, man-in-the-middle attacks, and protocol attacks. Some of these attacks have already been discussed in the previous discussion. Traffic analysis attacks are passive attacks where the attackers passively listen to the traffic and try to make sense out of it. These attacks are very hard to mitigate because the communicating parties usually have no idea that their traffic is being monitored. The attackers look for interesting information in the internet traffic such as users' personal information, business logic details, credentials, and other information that is of any value to the attacker. Besides, data transfer security is also of paramount importance in IoT. The data produced in the IoT environment is used for decision-making purpose; therefore, it is essential to guarantee the quality and health of the data. The compromise in the security of the data in IoT will have dire consequences on the underlying applications. 
It is worth mentioning that SDN has been leveraged to achieve a range of benefits for IoT applications and IoT security \cite{Szymanski2017}. 
The rich functionality provided by SDN control plane enables organizations to efficiently control millions of sensors and things in the IoT paradigm. However, despite all the aforementioned advantages of SDN, the open interfaces in SDN open pandora of new attacks on already vulnerable IoT devices, infrastructure, and applications \cite{Hayward2013, Dacier2017}. Therefore, the security of IoT is directly dependent on the security of SDN as well. 

\subsection{Security Issues in Cloud-Based IoT}
Another very important enabler for IoT is the cloud computing which is leveraged for processing massive data generated by the IoT subsystems. Security consideration in IoT is of prime importance from cloud tenants perspective as well as from service providers perspective \cite{Singh2016}. It is worth mentioning that cloud can be an essential part of the IoT infrastructure due to several-fold reasons: handling big data, storing and processing huge amount of data from IoT, and producing end-results to the respective applications in IoT environment. Additionally, cloud platform also provides a range of services such as device management, resource management, data processing, analysis, and management. The sheer scale of IoT devices pose a serious challenge for achieving the required security and privacy goals for IoT. Furthermore, the IoT applications are designed from a single domain in mind and thus-forth do not encompass the whole range of other domains that may use the data originated from one particular IoT domain. Therefore, the number of devices added to IoT networks and the data produced by these devices, and then stored processed, and analyzed by the cloud need viable, efficient, and scalable security and privacy considerations. 

\section{IoT Security and Machine Learning} 
\label{sec:MLDL}
In this section, we discuss various machine learning algorithms and their applicability in IoT applications.

\subsection{Basic Machine Learning Algorithms}

\begin{figure*}
\centering
\includegraphics[width=1\textwidth]{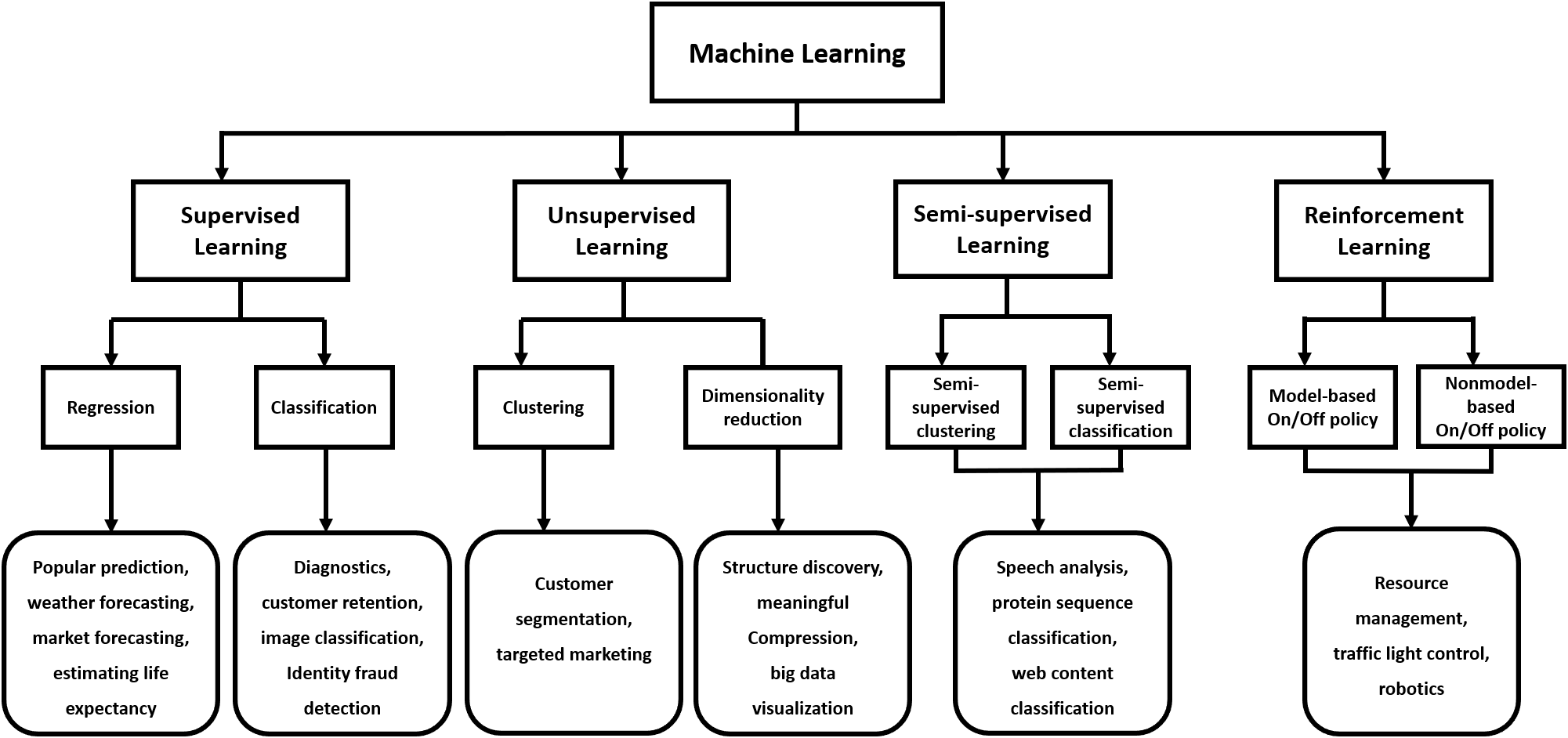}
\caption{Machine learning classes.}
\label{class}
\end{figure*}

The ML algorithms can be classified into four categories; supervised, unsupervised, semi-supervised, and reinforcement learning algorithms (Fig. \ref{class}). 

\noindent
\vspace{0.1cm}
{\textbf{Supervised Learning}}:
Supervised learning is performed when specific targets are defined  to reach from certain set of inputs. For this type of learning, the data is first labeled followed by training with labeled data (having inputs and desired outputs). It tries to identify automatically rules from available datasets and define various classes, and finally predict the belonging of elements (objects, individuals, and criteria) to a given class.

\noindent
\vspace{0.1cm}
{\textbf{Unsupervised Learning}}:
 In unsupervised learning, the environment only provides inputs without desired targets. It does not require labeled data and can investigate similarity among unlabeled data and classify the data into different groups. 
 
 \vspace{0.1cm}
Supervised learning and unsupervised techniques mainly focus on data analysis problems while reinforcement learning is preferred for comparison and decision-making problems. This categorization and choice of ML technique depends on the nature of available data. When the type of input data and the desired outputs (labels) are known, supervised learning is used. In this situation, the system is only trained to map inputs to the desired outputs. Classification and regression are the examples of supervised learning techniques where regression works with continuous and classification works with discrete outputs. Various regression techniques such as Support Vector Regression (SVR), linear regression, and polynomial regression are commonly used techniques. On the other hand, classification works with discrete output values (class labels). Common examples of classification algorithms include $K$-nearest neighbor, logistic regression, and Support Vector Machine (SVM). Some algorithms can be used for both classification and regression such as neural networks. When outputs are not well-defined and the system has to discover the structure within the raw data, unsupervised learning methods are used to train the system. Unsupervised learning includes clustering which groups objects based on established similarity criteria such as $K$-means clustering. The degree of precision of the predictive analytics depends on how well the respective ML technique has used past data to develop models and, how well is it able to predict the future values. Algorithms such as SVR, neural networks, and Naive Bayes are used for predictive modeling.

\noindent
\vspace{0.1cm}
{\textbf{Semi-supervised Learning}}:
In the previous two types, either there are no labels for all the observation in the dataset or labels are present for all the observations. Semi-supervised learning falls in between these two. In many practical situations, the cost to label is quite high, since it requires skilled human experts to do that. So, in the absence of labels in the majority of the observations but present in few, semi-supervised algorithms are the best candidates for the model building. 

\noindent
\vspace{0.1cm}
{\textbf{Reinforcement Learning}}:
In Reinforcement Learning (RL), no specific outcomes are defined, and the agent learns from feedback after interacting with the environment. It performs some actions and makes decisions on the basis of the reward obtained. An agent  can be rewarded for performing good actions or punishment for bad actions and use feedback criteria to maximize the long term rewards. 
It is greatly inspired by learning behaviors of humans and animals. Such behaviors make it an attractive approach in highly dynamic applications of robotics in which the system learns to accomplish certain tasks without explicit programming \cite{Ambedkar}. It is also very important to choose the suitable reward function because the success and failure of the agent depends on the accumulated total reward\cite{Wirth}. 



\subsection{Deep Learning (DL) and Deep Reinforcement Learning (DRL)} 

\noindent
\vspace{0.1cm}
{\textbf{Deep Learning}}:
DL is a machine learning technique originated from ANN. The neural network is comprised of neurons (considered as variables) connected through weighted connections (considered as parameters). To achieve the desired set of outputs, supervised or unsupervised learning technique is associated with the network. The learning is carried out by using labeled and unlabeled data from supervised or unsupervised learning techniques, respectively followed by the iterative adjustment of the weights among each pair of neurons. Therefore, while discussing about DL, we refer to large deep neural networks where the term deep refers to the number of layers in that network \cite{Xiao}, \cite{Lane}.


DL is known for distributed computing and, learning and analysis of sheer amount of unlabeled, un-categorized, and unsupervised data. It develops a hierarchical model of learning and feature representation motivated by layered learning process in the human brain \cite{Wang}.  DL models contribute to various ML applications such as speech recognition, computer vision and NLP by providing improved classification modeling and generating better data samples. Furthermore, these models also benefit the data compression and recovery, both in time and spatial domains because of its effectiveness in extracting patterns and features from large amounts of data and extracting relationships within time-dependent data.  The different DL architectures available in literature include, Convolutional Neural Networks (CNN), Recurrent Neural Networks (RNN), Boltzmann Machine (BM),  Long-short Term Memory (LSTM) Networks, Generative Adversarial Networks (GAN), Feed forward Deep Networks (FDN) and  Deep Belief Networks (DBN).  CNN (used in spatially distributed data) and RNN (used in time series data) are the most widely used deep learning architectures. 



\noindent
\vspace{0.1cm}
{\textbf{Deep Reinforcement  Learning}}: 
DL is one type of ML techniques used for function approximation, classification, and prediction whereas RL is another type of ML techniques used for decision
making in which a software agent learns about optimal actions by interacting with an environment over various states.  DL and RL come into play together in situation when the number of states and data dimensionality are very large and the environment is non-stationary. Therefore, traditional RL is not efficient enough. By combining DL and RL,  agents can learn by themselves and come up with a good policy to obtain maximum long-term rewards. In this approach, RL obtains help from DL to find the best policy and DL performs action values approximation in order to find the quality of an action in a given state. Furthermore, RL and DL benefit from each other.  DL is capable of learning from complex patterns but is prone to mis-classification. In this situation, RL has a powerful capability to automatically learn from environment without any feature crafting and helps DL in efficient classification \cite{Mufti,Mehdi,Ngoc}. DRL integrates RL's decision making and DL's perception. This combination has been used in game playing program ``AlphaGo" developed by Google \cite{Dongxia}. It can help solving tasks with high dimensional and raw data with some policy controls \cite{Sandeep}. Most recent work in DRL can be found in \cite{Kai,Lake}.

 
 \subsection{Machine Learning Techniques Used in IoT Security} 
 
In the following, we discuss various ML algorithms focusing on the underlying security and privacy problems in IoT networks, as shown in Table \ref{table:ML technique}. 
More precisely, we consider authentication, attack detection and mitigation, Distributed Denial of Service (DDoS) attacks, anomaly and intrusion detection, and malware analysis.

 \begin{table*}
\caption{ Machine Learning Techniques Used in IoT Security}
\label{table:ML technique}
\begin{center}
\begin{tabular}{|p{4cm}|p{8cm}|}
\hline
{\bf Machine Learning Algorithm} & {\bf Description}\\
  \hline
  \hline
  NaiveBayes & It is the classification algorithm used with binary and multi-class environment. It is named as ``Naive", as over-simplified assumptions are made for the calculation of probabilities for specific hypothesis. All the attributes are assumed to be conditionally independent instead of calculating the actual values \cite{Deng2018}. \\
   \hline
  $K$-Nearest Neighbour & It is simple and effective supervised learning algorithm and is used for associating new data points to the existing similar points by searching through the available dataset. The model is trained and grouped according to some criteria and incoming data is checked for similarity within $K$ neighbours \cite{Doshi2018}.\\ 
  \hline
    $K$-Means Algorithm & The most commonly used well know technique is $K$-means clustering algorithm belonging to the unsupervised category of ML family.  $K$-Means clustering is used to classify or group devices based on attributes or parameters, into $K$ number of groups, where $K$ is a positive integer number and 
  its value has to be known for the algorithm to work  \cite{AlaganF1}.\\ 
 \hline
      Random Forest and Decision Tree (DT) & It is a supervised learning method. It defines a model by implementing certain rules inferring from the data features. Afterwards, this model is used to predict the value of new targeted variable. Decision tree is used in classification and as well as regression problems. Essentially, these trees are used to split dataset into several branches based on certain rules \cite{Alam2013}. \\
      \hline
       Support Vector Machines (SVM) & SVM is a supervised ML algorithm with low computational complexity, used for classification and regression. It has the ability to work with binary as well as with multi-class environments \cite{Zhou2018},\cite{Ham2014}.
It classifies input data into $n$ dimensional space and draws $n-1$ hyperplane to divide the entire data points into groups. \\ 
      \hline
      Recurrent Neural Networks (RNN) & This is a supervised learning algorithm used to develop a cascaded chain of decision units for solving the complex problems \cite{Chauhan2018}. It essentially constructs network with certain number of inputs to trigger outputs. Various types of neural networks have been proposed in the literature, e.g. Multi-Layer Perceptron (MLP), Convolutional Neural Networks (CNNs), and Recurrent Neural Networks (RNNs) \cite{HADDADPAJOUH2018},\cite{KARBAB2018},\cite{Su2018}.\\
      \hline
      Principal Component Analysis &  It is an unsupervised ML algorithm and is multivariate technique for data compression. It performs dimensionality reduction in large data sets and extracts useful information in the form of set comprised of orthogonal variables known as ``principal components". These components are organized in an increasing order of variance where first component is associated with highest variance of the data and it continues to the last. The least variance components having least information can be discarded \cite{An2017}.\\
\hline
Q-Learning  & It is used for resource scheduling in spectrum management in addition to security in IoT. Q-learning belongs to reinforcement learning (RL) class of the ML. In RL, an agent learns by trial and error that how its actions effect the environment. It estimates the reward after each action and moves to the new state accordingly \cite{AlaganF1}. It will get reward for good action and penalty for bad actions.\\
\hline
Deep Learning & It is essentially a feed forward Neural Network (NN) in which each neuron is connected to another layer and no connection exists within the layer. The term deep learning refers to multiple layers holding multiple levels of perception such that each layer receives input from the previous layer and feeds the result to following layer \cite{Azmoodeh2018}.\\
\hline
  \end{tabular}
\end{center}
\end{table*}

\begin{table*}
\caption{Security Problems in IoT Networks and Applied Machine Learning Techniques}
\label{table:research objective}
\begin{center}
\begin{tabular}{|p{5cm}|p{7cm}|}
\hline
{\bf Research Objective/Problems} & {\bf Machine Learning Techniques(Surveyed References)}\\
  \hline
  \hline
    Authentication & \begin{itemize}
      \item Deep Learning
      \item Recurrent neural networks (RNNs) \cite{Chauhan2018}
      \item Q-learning and Dyna-Q \cite{Xiao2016}
      \item Deep Neural Network (DNN) \cite{Shi2017}
     \end{itemize}\\
  \hline
  Attack Detection and Mitigation &\begin{itemize}
  \item SVM 
      \item Deep Learning \cite{DIRO2018},\cite{Abeshu2018},\cite{RATHORE2018}
      \item Unsupervised learning, stacked autoencoders
      \item Extreme Learning Machine (ELM)-based semi-supervised Fuzzy C-Means (ESFCM)
      \item $K$-Nearest Neighbour (NN) and SVM \cite{Ozay2016}
      \end{itemize}\\
  \hline
 Distributed DOS Attack &\begin{itemize}
          \item $K$-Nearest Neighbour \cite{Doshi2018} 
          \item Support Vector Machine \cite{Doshi2018}
          \item Random Forest and Decision Tree \cite{Doshi2018}
          \item Neural Network \cite{Doshi2018} 
          \item Multivariate Correlation Analysis (MCA) \cite{Tan2014}
          \item Q learning \cite{Li2017}
      \end{itemize}\\
  \hline
  Anomaly/Intrusion Detection & \begin{itemize}
      \item $K$-means clustering and Decision Tree  \cite{Shukla2017}
      \item Artificial Neural Network ANN \cite{Canedo2016}
        \item Novelty and Outlier Detection \cite{NESA2018}
        \item Decision Tree \cite{VIEGAS2018}
        \item Naive Bayes \cite{VIEGAS2018},  \cite{Haddad2018}
        \end{itemize}\\
  \hline
 Malware Analysis & \begin{itemize}
          \item Recurrent Neural Network (RNN) \cite{HADDADPAJOUH2018}
          \item ensemble learning algorithm Random Forest supervised classifier \cite{Alam2013}
          \item Deep Eigensapce Learning and Deep Convolutional Networks \cite{Azmoodeh2018}
          \item SVM \cite{Zhou2018}
          \item PCA, one-class SVM, and naive anomaly detector based on unseen $n$-grams \cite{An2017}
          \item CNN \cite{Su2018}
          \item Artificial Neural Network
          \cite{KARBAB2018}
          \item Linear SVM \cite{Ham2014}
          \item SVM and PCA \cite{Esmalifalak2013,Esmalifalak2017}
  \end{itemize}  \\
  \hline
  \end{tabular}
\end{center}
\end{table*}

Supervised learning algorithms work with labelled data and are utilized in IoT networks for spectrum sensing, channel estimation, adaptive filtering, security, and localization problems. This category holds two distinct types of techniques: classification and regression. Classification under supervised machine learning is used for predication as well as modeling of the available data sets. Regression is used for predicting continuous numeric variables. SVM, Naive Bayes, Random Forest, Decision Tree are few of the widely used classification algorithms. SVM uses a mechanism called kernels that is used to find difference between two points of the two separate classes. SVMs are able to model non-linear decision boundaries.  However, SVM is inherently memory intensive and is difficult to decide suitable kernel, and it becomes difficult to model large datasets. Therefore, random forests are usually preferred over SVM. While Naive Bayes (NB) is used to model real world problems like text classification and spam detection. Being naive and all input features being independent of each other makes random forest algorithms ideal for modelling real world problems.  Random forest algorithms are easier to implement and are adaptive to the size of available data set. These algorithms take longer time to train as compared to other supervised algorithms such as SVM and NB. But it achieves higher degree of accuracy and takes less time for prediction. Furthermore, it is based on constructing the graph with branches and leafs, signifying decision and class respectively. For classification of an event, a top-down approach is used by traversing the tree until a decision is made for a class.  Nearest neighbors and logistic regression are to famous regression algorithms. These algorithms are also known as ``instance-based", that make predictions for each new observation by searching for the most similar training data. However, these algorithms are memory-intensive and perform poorly for high-dimensional data. 

The family of unsupervised learning algorithms deals with unlabeled data and utilize input data in a heuristic manner. These are used in anomaly, fault, and intrusion detection, cell clustering, and load balancing. Clustering under unsupervised learning category is used for data groupings based on some inherent similarities and dissimilarities. The clustering is unsupervised, and therefore, there are no right or wrong answers. To evaluate the accuracy of the results, data visualization is used. If there is a possible right or wrong answer, then the clusters can be pre-labeled in datasets and in this scenario, classification  algorithms are preferred. $K$-means and hierarchical clustering are two popular clustering algorithms. $K$-Means is most popular because it is a simple and flexible algorithm that forms clusters based on geometric distances between data points. Clusters are grouped around centroids resulting into globular with the same size. However, the  of clusters has to be specified before clustering starts and it is not always possible and efficient to do.  Also, if clusters are not globular, it results into poor cluster formation. Most of the IoT applications, falls under the unsupervised  learning approaches with very less initial information about the environment, similar to natural learning processes of the living species. For instance, the zero-day attacks on IoT networks have little or no information to start with. 

RL techniques learn by exploiting various stages and develop the reward and action relationship between agent and the environment. This relationship of action reward is very useful in solving various IoT problems \cite{Walid}.  It does not require extensive training data set; however, the agent is required to have the knowledge of the state transition function. It is computationally simple but requires significant time to converge to a steady state. This slow convergence and knowledge of the state transition function or optimal policy are the key challenges in using RL algorithms in  dynamic environments of IoT networks. 

DL relies on strong function approximation, estimation and the learning capabilities thus providing more efficient solutions in various problem areas of IoT domain and security/privacy is no exception. IoT devices due to their resource constraints may not be able to host or run complex computational algorithms for any type of task such as communication, analytic and prediction. Therefore, DL-based algorithms show better performance with lower latency and complexity compared to conventional theories and techniques \cite{Wang}. Additionally, DNNs are good in locating and defining low dimensional representations from any type (text, image, audio) of high dimensional data patterns. DRL and its variants are used for authentication and DDoS detection in heterogeneous IoT networks. The major DRL algorithms used for security and privacy are; deep deterministic policy gradient, continuous DQN, prioritized experience replay, asynchronous
$N$-step Q-learning,  deep SARSA, Dueling network DQN, and asynchronous advantage actorcritic.  

Table \ref{table:research objective} lists different security challenges along with utilized ML and DL techniques.

\subsection{Limitations in Applying Machine Learning in IoT Networks} 
IoT traffic is usually characterized by its sheer volume, diversity, variable speed, and uncertainty. 
Most of the traditional ML techniques are not inherently efficient and scalable enough to manage IoT data and thus need considerable modifications \cite{Junfei}.  Moreover, inherent uncertainties exist in IoT data and is difficult to expunge these intrinsic unpredictability. In the following, we discuss some of the common limitations of using ML techniques in IoT networks.

\subsubsection{Processing power and energy}  
Machine learning algorithms inherently possess some complexity issues such as memory, computational, and sample complexity. Also, conventional ML approaches lack scalability and are only limited to low-dimensional problems. IoT devices are small and typically have energy constraints with limited processing power. Therefore, direct application of conventional ML techniques is not suitable in resource-constrained environments. On the other hand, smart IoT devices require real-time data processing for real time applications, while traditional ML techniques are not designed to handle constant streams of data in a real-time. In the wake of such limitations, it is imperative to merge the existing streaming solutions with ML algorithms; however, it will increase the overall complexity of an algorithm. 

In addition to this, ML-based networks are developed assuming that the entire data set is available for processing during training phase. However, this is not true for the IoT data. This phenomenon gives rise to various challenges when traditional techniques have to handle an unprecedented volume of data. Also, predictive ability of an algorithm decreases with the increase in the dimensionality of data \cite{Alexandra}. The preceding discussion is, at par, applicable for the security-related functions in the IoT where real-time data is processed for possible attack vectors such as intrusion and so on.  

\subsubsection{Data management and analytics} 
Wireless data can be generated from different sources including networked information systems, and sensing and communication devices \cite{Tadilo}. Data is the crown-jewel for IoT systems where efficient analysis must be performed to obtain meaningful information from the data; however, massive data management is a serious challenge in IoT from every application standpoint. The data generated in IoT networks is diverse in nature with different types, formats and semantics, thus exhibiting syntactic and semantic heterogeneity. Syntactic heterogeneity refers to diversity in the data types, file formats, encoding schemes, and data models. While semantic heterogeneity refers to differences in the meanings and interpretations of the data. Such heterogeneity leads to problems in terms of  efficient and unified generalization, specifically in case of big data and various datasets with different attributes. 

Machine learning assumes that the statistical properties across the entire dataset remain the same, and requires pre-processing and cleaning of the data before fitting within a specific model. However, that is not the case in real-world where data from various sources have different formatting and representations. Furthermore, there might also be differences among different parts of the same dataset. This situation causes difficulties for machine learning algorithms because the algorithms are usually not designed to handle  semantically and syntactically diverse data. This phenomenon advocates for efficient solutions to the heterogeneity problem.

\section{Survey of Existing Machine Learning-based Solutions for IoT Security}
\label{sec:MLSecSolutions}
In this section, we survey the existing ML-based solutions addressing different security issues in IoT.

\subsection{Authentication and Access Control in IoT}
Authentication is one of the primary security requirements in IoT. The users must be authenticated in order to use IoT applications and/or services. Typically, IoT applications and services are based on data exchange across different platforms. The data retrieved from the IoT devices is pre-processed, processed, and then passed through a decision-support system to make sense out of it. These processes may vary depending upon the underlying IoT architecture; however, data flow may be identical in these systems. Without loss of generality, when an application and/or a user needs some data from an IoT device, the entity (user or application) must be authenticated to IoT network and it should be made sure that the requester has required access rights for the data. Otherwise, the request to access such data will be denied. Like other networks, access control is also of paramount importance in IoT networks and equally challenging due to, but not limited to, network heterogeneity, volume of network, resource constraints of the devices, network (in)security, and attacks vulnerabilities, to name a few. Furthermore, it is also important to grant and revoke the access of certain users to the critical data of IoT applications and services. Before, we discuss ML-based access control mechanisms in IoT, it is important to put light on different categories of access control. 

Taylor et al. divided the access control mechanisms in IoT into three categories, Role Based Access Control (RBAC), Context Aware Access Control (CWAC) and Policy Based Access Control (PBAC) \cite{TAYLOR2017}. Whereas Ouaddah et al. extended this classification into more categories that include Attribute-based Access Control (ABAC), Usage Control-based Access Control (UCAC), Capability-based Access Control (CAC), and Organizational-based Access Control (OAC) \cite{OUADDAH2017}.  Most of the existing work fall under one of these categories. Ouaddah et al. conducted a comprehensive survey on the currently used access control mechanisms in IoT and identified challenges faced by access control in IoT \cite{OUADDAH2017}. The authors discuss the existing mechanisms based on the underlying IoT applications. IoT applications can broadly be divided into two major classes, personal and enterprise. Personal applications include smart homes, health-care, smart office, body area networks, sensors networks, whereas enterprise applications include smart cities, smart industries, critical infrastructure, and so on. Access control mechanism can be used both at application layer and at the architecture layer. At the architecture layer, the service providers have variety of choices such as defining an access control markup language such as Extensible Access Control Markup Language (XACML) \cite{Aziz2013,Diez2017}, Open Authorization (OAuth) \cite{Cirani2015}, and User-Managed Access (UMA) \cite{Rivera2015}. These architectural level access control mechanisms are already being used by many existing protocols of IoT, for instance OAuth is implemented over existing IoT protocols such as MQTT\footnote{\url{http://mqtt.org/}} and Constrained Application Protocol (CoAP)\footnote{\url{http://coap.technology/}}. Furthermore, services providers such as Google, Facebook, and Microsoft, Instagram, Flikr, Netflix, and many more, have billions of user accounts that use OAuth. Next, we discuss the existing ML- and DL-based authentication and access control mechanisms in IoT networks.

\subsubsection{ML-based authentication and access control in IoT}
 Xiao et al. \cite{Xiao2016} proposed a physical layer authentication mechanism for IoT. The proposed authentication mechanism uses physical channel properties such as signal strength. In essence, Xiao et al. used two revolutionary methods for physical layer authentication, game theoretic approach and machine learning technique, to isolate spoofers from benign IoT users. The authentication mechanism is formulated as a zero-sum game where the spoofing nodes try to increase their utility through maximizing the attack frequency. Whereas the channel frequency responses are used to establish Nash Equilibrium (NE).  It is worth mentioning that the packets received through radio interface have certain channel state which could be leveraged to test for a particular threshold based on which authentication decision is taken. For this purpose the authors used reinforcement learning (Q learning) and Dyna-Q which helps in understanding the channel state without having detailed information about the channel. The authors through experiments concluded that the detection accuracy and performance of Dyna-Q is better than Q-learning technique.  Another similar physical-layer authentication scheme based on distributed Frank-Wolf (dFW) algorithm is proposed in \cite{Xiao2017}. 
 
 \subsubsection{DL-based authentication and access control in IoT}
 Shi et al. \cite{Shi2017} proposed a user authentication technique for IoT based on human physiological activities through WiFi signals. The proposed authentication scheme is based on both activity recognition and human identification. Activity recognition can be performed with coarse-grained data and a smaller number of features. For this purpose, the channel state information in WiFi signals generated by IoT devices is used to derive characteristics of different features of the activities. Shi et al. use 3 layer Deep Neural Network (DNN) to learn the human physiological and behavioral characteristics that are used in authentication. In the three layers, the DNN extracts the type of activity (whether walking or stationary), then at the second layer, details of the activity are learnt, and at the third layer, high-level features based on which, the user is authenticated.

\subsection{Attack Detection and Mitigation}
In the preceding subsections, we described in detail, different attacks launched on IoT at different layers. The heterogeneity and scarcity of resources in IoT devices make them a perfect playground for attackers. Usually attackers exploit the known vulnerabilities in both networks and in the device and launch different kinds of attacks. It is worth mentioning that these attacks vary from a low-profile hacking into a device for fun to a massive scale ransomware attacks such as WanaCry and even more sophisticated attacks such as Mirai and Dyn. The traditional attack detection and mitigation mechanisms are based on cryptographic primitives and sometimes suffer from lack in accuracy and cause false positives. Therefore, ML-based techniques such as SVM, DL, autoencoders, $K$-NN, unsupervised learning, and so on. Here we briefly discuss the ML-based methods leveraged for attack detection and mitigation in IoT networks. In the following we investigate the existing ML-and DL-based techniques 

\subsubsection{ML-based attack detection and mitigation in IoT}
Rathore et al. \cite{RATHORE2018} proposed a semi-supervised learning-based attack detection mechanism for IoT. In essence, the proposed scheme is based on Extreme Learning Machine (ELM) algorithm and leverages Fuzzy C-Means (FCM) methods, collectively referred to as ESFCM. ESFCM is also implemented in fog infrastructure. One distinct feature of ESFCM is that it deals with labeled data, therefore it increases the detection rate of distributed attacks. However the detection accuracy of ESFCM is less than the previous two DL-based mechanisms but it outperforms the traditional ML algorithms for attack detection. Nevertheless, the semi-supervised learning mechanism harnesses the features of both supervised and unsupervised learning and makes it more efficient than its counterparts.

As aforementioned, IoT has many breeds ranging from personal networks (body area network) to more sophisticated critical industry infrastructure such as smart grid. Attack detection is important at par in these infrastructures. For instance in smart grid, the measurements are critical and must be genuinely retrieved and not tampered with as a result of an attack.  In this direction, Ozay et al. \cite{Ozay2016} studied in detail, different ML algorithms for attack detection in smart grid. The authors particularly investigated the role of supervised learning, semi-supervised learning, feature space fusion, and online learning algorithms for attack detection. The authors divided the networks into small and large networks and found out through experiments that in small networks, $k$-NN performs better whereas in large networks, SVM performs better in terms of the attack detection accuracy.

\subsubsection{DL-based attack detection and mitigation in IoT}
Diro et al. \cite{DIRO2018} proposed a DL-based attack detection mechanism in IoT by leveraging fog ecosystem. In essence, the attack detection mechanism is implemented at the edge near the smart infrastructure. The distributed attack detection mechanism takes into account different parameters from the learning mechanism and decides upon the output of the learning architecture on a given data.  
The rationale for using fog infrastructure is the resource constraints and the application nature of IoT. In case of critical infrastructure the learning mechanism must be as near as possible to the data-generating nodes in order to take timely and informed decision in case of potential attacks. Similarly Abeshu et al. \cite{Abeshu2018} proposed a distributed DL-based attack detection mechanism in IoT. They used fog computing architecture (which is one of the favored architecture to realize IoT applications) to implement DL techniques for attack detection. The proposed technique focused on fog-to-thing communication where the learning module is implement at the fog layer which is the optimal point for detection mechanism because it both reduces the latency for communication and utilizes the resources.

\subsection{DoS and Distributed DoS (DDoS) Attacks}

DoS and DDoS attacks are two of the most notorious attacks and hard to mitigate in IoT environment. There are several-fold reasons that make it hard to provide effective solutions against such attacks. These reasons include the sheer number of connected IoT devices to the Internet, heterogeneity, poor security mechanisms due to resource constraints of the IoT devices, cross-platform communications, massive scale communication, and so on. These phenomena bring the vulnerabilities of IoT devices against DDoS attacks to the fore. Such attacks must be prevented to guarantee the smooth functionalities of IoT applications. In this regard, Vlajic et al. \cite{Vlajic2018} referred to the IoT as the "Land of opportunities for DDoS attackers". 

The year 2016 witnessed an unprecedented growth in attacks against IoT infrastructure with a massive scale. Mirai\footnote{\url{https://www.csoonline.com/article/3258748/security/the-mirai-botnet-explained-how-teen-scammers-and-cctv-cameras-almost-\\brought-down-the-internet.html}} was one of these attacks that almost brought down the Internet where household devices such as babycams, printers, and webcams etc. were used as bots to launch DDoS attacks on many organizations. Similarly, other Mirai-like bots have been reported in \cite{Kolias2017}. Since its inception in 2016, the Mirai is a family of malware that caused serious disruptions in Internet services due to its sophisticated spreading mechanism in the IoT network. A detailed surgery of Mirai malware can also be found in \cite{Kambourakis2017}.  The fact that IoT devices are not only hacked but used as frontline tools to launch notorious attacks against large organizations, advocates for serious intelligent security measures that will secure these devices. To date, noteworthy research results have been yielded through different mechanisms to mitigate DDoS attacks in IoT; however, different architectures make it really hard to devise a unified mechanism to combat DDoS attacks in different IoT platforms \cite{Yin2018}.  

Traditional DDoS detection and mitigation mechanisms in IoT networks are applied at the gateways, routers, and the entry points for IoT networks with the help of both intrusion detection and prevention mechanisms. As mentioned before, MQTT and CoAP are two of the most widely used telemetry protocols in IoT.  Pacheco et al. \cite{Pacheco2016} evaluated the effect of DDoS attack on generic IoT network that uses CoAP. This work used only reflection attack \cite{Paxson2001} to assess the withstanding capabilities of IoT network. In addition to detection and mitigating DDoS attacks in IoT from inside, other enabling technologies such as fog and cloud computing have also been used to aid the DDoS detection mechanism in IoT. For instance, Alharbi et al. \cite{Alharbi2017} used fog computing based approach to secure IoT communications and mitigate malicious attacks. 

Among other flavors of IoT networks, industrial and critical infrastructure-aided IoT networks must exhibit strong resilience to DDoS attacks. Similarly, a fog and cloud computing-based DDoS mitigation framework is proposed in \cite{Yan2018}. In this multi-level DDoS detection framework, traditional mechanisms are used at multiple layers of industrial IoT infrastructure. The lowest level uses edge computing with SDN gatweways, and then network traffic data is gathered at the fog computing level which consists of SDN controllers, and analyze the data for possible DDoS. Furthermore, honeypots are also utilized at this level. Finally at the cloud computing level, applications-generated data is analyzed at the cloud platform to detect any threat of potential DDoS attack. 

From the preceding discussion, it is clear that there is no silver bullet that could solve the DDoS problem in resource-constrained networks and it is also clear from literature that more intelligent mechanism is essential to detect and mitigate DDoS. Furthermore, false positives are still not out of question in such cases and benign requests might be blocked. Therefore, the resource-constrained and poor security-enabled devices in IoT networks provide alluring playground for the malicious attackers. Despite the recent advancements in mitigating such types of attacks, it is still essential to work on intelligent mechanisms that not only take into account the amount of traffic, but also the behavior of the attackers. In this context, machine learning is the most suitable candidate to be leveraged for DDoS detection in IoT networks. To mitigate DDoS attacks in IoT networks, a number of ML and DL techniquest have been leveraged by the research community. 

\subsubsection{ML-based techniques to address DoS and DDoS attacks in IoT}
Doshi et al. carried out a detailed comparison of the existing ML mechanisms for DDoS detection in IoT networks \cite{Doshi2018}. They leveraged the distinctive features of the IoT traffic where IoT devices usually engage in a finite communication with end-points rather than back-end servers. Doshi et al. compared K-nearest neighbors, decision trees, neural networks, random forest, and SVM to detect DDoS in IoT. Similarly, Ye et al. \cite{Ye2018Hindawi} proposed SVM-based DDoS detection mechanism in SDN environment which complements the IoT applications.  
Another such SVM-based DDoS detection in SDN is proposed by Kolika et al. \cite{Kokila2014} with almost same accuracy as that of Ye et al.'s scheme \cite{Ye2018Hindawi}. Kolika et al. also compared other techniques such as Naive Bayes, Random Forest, Bagging, and Radial Basis Function (RBF).

Besides SVM, other ML techniques have also been used for DDoS detection. For instance, Tan et al. \cite{Tan2014} leveraged Multivariate Correlation Analysis (MCA) for DDoS detection. MCA-based DDoS detection mechanism focuses on the server side and in the context of the IoT, this mechanism will detect DDoS attack as a result of data flow between the back-end servers for data gathering, processing, and decision-making. In essence, MCA techniques is based on behavioral analysis of the traffic where normal behavior is isolated from abnormal. Features are generated from network traffic and then MCA is applied on those features. MCA determines and extract the correlation between the features obtained from the first step. Ideally, the intrusions cause disruption in the correlation among features which could be an indication for possible intrusive activity. Finally, any known anomaly detection technique is used for decision making.  

A Signal-to-Interference-plus-Noise-Ratio (SINR)-based DoS attack detection mechanism for Cyber-Physical System (CPS) is proposed by Li et al. \cite{Li2017}. The authors have formulated the DoS attack as a game between the sensor and attacker with multiple energy levels. Furthermore, in this work, the authors focus on the transmission power consumption for the sensor and the interference power consumption for the attacker. To establish a balanced equilibrium between the players of the game, the authors use Nash Q-learning algorithm.

\subsubsection{DL-based Techniques to Address DoS and DDoS Attacks in IoT}
Hodo et al. \cite{Hodo2016} used supervised Artificial Neural Networks (ANNs) technique to thwart DDoS in IoT.  
Similarly, Kulkarni et al. \cite{Kulkarni2009} used Multi-Layer Perception (MLP) mechanism to detect DoS attack in sensor networks which is mainly used in IoT.

\subsection{Anomaly/Intrusion Detection}
Here we outline briefly, the existing technique to both detect and mitigate intrusion in IoT. We briefly outline the common techniques for intrusion detection and focus in detail on the ML-based intrusion detection mechanisms in IoT. To date, a number of ML-based techniques have been used to detect anomalies and intrusions in the IoT networks and their different breeds \cite{Branch2013,Alsheikh2014}. Traffic filtering is one of the most widely used mechanism for intrusion detection where per packet analysis or batch analysis is performed to isolate legitimate packets from the malicious ones. However, despite the effectiveness of traffic classification, the higher number of false positives produced as a result of such classification, render this method to be less reliable. On the other hand, behavior-based models are also used to detect intrusion in the network. In the context of IoT, both traffic classification, and behavior-based models are used. Meng et al. \cite{Meng2018} discussed a trust-based approach for intrusion detection in IoT networks. Unlike the traditional mechanisms, trust-based approach takes into account the level of confidence in a device in combination with the type and class of traffic. The authors proposed the combination of trust management and traffic classification to detect intrusion in the IoT networks. 

However, it is important to note that traditional signature-based and behavior-based schemes fail to detect zero-day intrusions. Therefore, artificial intelligence and its breeds are employed in intrusion detection system (IDS). Li et al. \cite{Li2018AIbasedTI} proposed an Artificial Intelligence- (AI) based mechanism for intrusion detection in SDN-driven IoT. This scheme is based on network traffic flow where the intrusion detection component of the network captures the flow and applies two algorithms for features extraction, i.e. Bat algorithm (with swarm division) and differential mutation. Afterwards random forest technique is used to classify the traffic flow to identify potential intrusion to the system. The aim is to improve the detection accuracy and reduce the false positives. Intrusion may occur in the IoT network through different types of attacks, the likes of which can be found in the normal networks such as spoofing, masquerading, distributed denial of service, hijacking control, and so forth \cite{Li2018}. 

In essence, IoT is the mixture of existing networking technologies that have paved their way into integration with our daily lives such that sensor networks, bluetooth, RFID, WiFi, and so on. Therefore, the security mechanisms, access control, and protection techniques are not homogeneous across different platform. In this context, the intrusion detection techniques need special attention depending on the underlying technologies. To address the issue of intrusion detection in different technologies, Gendreau et al. \cite{Gendreau2016} conducted a survey of intrusion detection techniques in different networking paradigms. Similarly, in \cite{Butun2014,Abduvaliye2013}, the authors focused on IDS in wireless sensors networks. As discussed earlier, a number of intrusion detection techniques are proposed in the literature to detect and/or mitigate network intrusion in IoT. Colom et al. \cite{COLOM201876} proposed a distributed framework for intrusion detection in IoT based on task distribution among different nodes depending on the security requirements and the state of available resources. A controller component in the proposed framework administers the intrusion detection in an attack scenario. 

Finally, an architecture-focused survey on IDS in IoT is carried out by Benkhelifa et al. \cite{Benkhelifa2018}. This survey highlights the existing protocols, detection methods, and points out the future research challenges in this direction. Intrusion detection in IoT is very hard to mitigate due to the inherent nature of the IoT and the fact that most of the intrusion detection methods are behavior-based, makes perfect sense to focus on the evolutionary methods of intrusion detection. In this spirit, ML and DL methods are employed for this purpose.

\subsubsection{ML-based IDS in IoT}

Shukla et al. \cite{Shukla2017} proposed ML-based lightweight IDS for low-power IoT networks running 6LoWPAN. They used the IDS mechanism to detect wormhole attacks in IoT networks. The proposed IDS mechanism uses three ML techniques, i.e. $K$-means clustering (unsupervised learning), decision tree (supervised learning), and a hybrid technique combining the aforementioned techniques.  
On the other hand, Canedo et al. \cite{Canedo2016} leveraged two ML techniques to detect intrusions at the IoT gateways. The authors used ANN and genetic algorithms for IoT security. 

Other ML techniques used for intrusion and anomaly detection include outlier detection, naive Bayes, RNN, decision tree, and DL. In \cite{NESA2018}, used outlier detection mechanism to deal with the unhealthy data in IoT networks. Traditional outlier detection methods include regression analysis, statistical methods that are known to require sheer amount of data to draw conclusions about the outliers in the data. Nesa et al. \cite{NESA2018} used non-parametric approach that is suitable for IoT because it does not require large storage to store the incoming data. The authors leverage sequence-based supervised learning which is based on Influential Relative Grade (IRG) and Relative Mass Function (RMF) which efficiently detects the outliers.  
Viegas et al. \cite{VIEGAS2018} aim at energy-efficient and hardware-friendly implementation of the intrusion detection systems in IoT. The authors leveraged 3 classifiers, Decision Tree (DT), Naive Bayes (NB), and Linear Discriminant Analysis (LDA) during their experimentation for intrusion detection. In this work, the authors first analyzed the effect of single classifier among the afore-mentioned classifiers and then used the combination of these classifiers to see the effect on the detection. 
Similarly, Sedjelmaci et al.  \cite{Sedjelmaci2016} also focused on the balance between energy consumption and intrusion detection in IoT networks. The authors used game theoretic approach for the detection of new types of intrusion in IoT networks.   

\subsubsection{DL-based IDS in IoT}

Deep learning is also leveraged for IDS in heterogeneous IoT networks. For instance, Recurrent Neural Network (RNN) is used by Kim et al. \cite{Kim2016} to train the IDS model which is based on Long Short Term Memory (LSTM) architecture.  
Similarly, Saeed et al. \cite{Saeed2016} used Random Neural Networks (RaNN) for the realization of efficient and fast anomaly-based intrusion detection in low-power IoT networks. The authors proposed a two-layer model where at the first layer, normal behavior is learnt by the system and at the second layer, different kinds of Illegal Memory Access (IMA) bugs and data integrity attacks on the network are detected. The proposed solution is centralized where the results are sent to a central server.  

\subsection{Malware Analysis in IoT}
The surge in the number and heterogeneity of IoT devices provide a perfect and lucrative playground for cyber-attackers. One of the most notorious attack domains is malicious code injection and execution in IoT devices by exploiting the existing vulnerabilities in IoT devices. The vulnerabilities that could be used for malware injection could be related to application security, authentication,  and authorization. Apart from these methods, tampering the IoT devices physically for software modification and misconfiguration of security parameters could also enable attackers for the injection of malicious code. Before diving into the details of the malware, it is important to understand the types of the malware that endanger the IoT security. A malware is a threat that persists as a result of the aforementioned vulnerabilities and executed through a number of attacks. The common types of malware include, but not limited to, bot, spyware, ransomware, adware, trojan, and virus, to name a few.  In this section, first we summarize the classification of malware that affect the IoT devices and then discuss the existing solutions including ML-based techniques that can keep IoT devices safe. 

It has been found out through many exploratory studies that there are enormous smart devices that are connected to internet without any proper security protection that do not only pose threats to the device itself, but also enable the attackers to tap resources for the attacks at massive scale such as DDoS. For instance Moos et al. \cite{Moos2017} tested his music device for vulnerability and exploited it to launch malicious-code attack which was very successful. Another project namely Insecam\footnote{\url{www.insecam.org}} lists the webcams that are potentially vulnerable to attacks and connected to Internet. These web-based cameras cover residential, public places, offices, and restaurants and can be used by attackers for malicious purposes. The types of Malware that have been successful in disrupting the normal functionality of the organization, application, or target entities include, but not limited to, NotPetya, Stuxnet, Cryptlocker, Red October, Night Dragon, and so on. 

Makhdoom et al. \cite{Makhdoom2018} provided a detailed taxonomy and working principles of these malwares. These are the generic malware attacks whereas there are optimized families of malware attacks that particularly target the IoT devices. Such attacks include WanaCry, Cryptlocker, Mirai, Stuxnet, and so on. These are the malware attacks that have costed the industry staggering amount of money and other loss such as public image of the company. The details of these attacks are out of the scope of this paper. However it is important to know the generalized approach of attackers to launch attack through malware \cite{Makhdoom2018}. First of all, the attackers gather knowledge about the potential target, for instance sensor networks, through reconnaissance. There are many methods to carry out reconnaissance, for instance tools such as Nmap, Metasploit, and Wireshark as well as social engineering. Some of the existing tools provide meta information about the exploits which is makes it even easier for the attackers.

The first step gives a clear idea to the attackers as to what kind of vulnerability to use against a specific class of devices. To date, standard application security assessment methods such as Open Web-Application Security Project (OWASP)\footnote{\url{https://www.owasp.org/
}} provides the major sources of vulnerabilities among which the attacker can choose appropriate vulnerability. In principle, OWASP provides the sources of exploitations that could be used by the attackers such as SQL injection, security misconfigurations, broken authentication, Cross-Site Scripting (XSS), and so on. Furthermore, depending on the type of the device and the exploit, the attackers can send the payload to the target through several means such as phishing, updates, rootkit, and so on. The malware families vary from a simple single-task malware to a more complex, intelligent, dormant and multi-purpose evolving malware.  
Today's intelligent malwares are also adaptive according to the IoT environment where they can assess the network, and adapt its execution according to the underlying network. For instance, some malwares can evade the detection mechanism and stay dormant for sometime and do not execute the malicious code until it is safe for them not to compromise their intent. In this regard, there are a number of countermeasure techniques for malware to evade the detection mechanisms \cite{You2010, Rudd2017, Veerappan2018}.  

\subsubsection{Malware evasion techniques}

You et al. \cite{You2010} outlined different approaches for malwares to evade the detection techniques. Encrypted malwares refer to an encrypted malicious code with respective decryptor to bypass the signature-based antivirus. However, since the decryptor remains the same for different versions of the same malware, therefore it can still be detected. To overcome this problem, the decryptor can be mutated and thus bypasses the detection mechanism. This type of malware is referred to as oligomorphic malware. Similarly, polymorphic malware generates countless decryptor which makes it even harder for detection engines to detect malware. Finally, metamorphic malware is the most sophisticated among the malware groups where it evolves to a new generation where it is different from the previous one and very hard to be detected. You et al. \cite{You2010} discussed the obfuscation techniques for polymorphic and metamorphic malwares that include dead-code insertion, register reassignment, instruction substitution, subroutine recording, and code integration. Furthermore, Rudd et al. \cite{Rudd2017} outlined in detail, the stealth malware and its mitigation techniques.
To date, a number of malware detection techniques have been proposed in the literature for different breeds of the IoT networks.

However, the traditional malware detection techniques might not be effective against sophisticated malwares; therefore, behavior-based intelligent malware analysis techniques are essential for IoT networks. In this context, ML and DL-based malware analysis techniques have been developed in the literature. Here we discuss these techniques in detail. ML and DL techniques such as RNN, random forest, Deep Eigenspace Learning (DEL), SVM, PCA, CNN, and ANN have been leveraged for malware analysis in IoT. 

\subsubsection{ML-based malware analysis in IoT}
Alam et al. \cite{Alam2013} used ensemble supervised learning technique with random forest classifier to detect android-based malware. The classifier is checked for the detection accuracy of the malware samples in android applications. In \cite{Zhou2018}, Zhou et al. investigated the malware detection and propagation in IoT-based Wireless Multimedia System (WMS). The authors proposed a cloud-based approach where SVM is leveraged to detect the potential malwares and their propagation, and used state-based differential game to suppress the malwares. After achieving the Nash Equilibrium, the authors try to find optimal strategies for WMS to defend against malwares. Similarly, Ham et al. \cite{Ham2014} proposed a linear SVM-based technique for malware classification in android-based IoT. The authors also compared the results with other classifiers. Although SVM incurs more classification time due to the removal of unnecessary features; however, it is favorable due to its less complexity and better accuracy. To assess the detection accuracy of the detection model, the authors considered different types of malwares and their features.  

An et al. \cite{An2017} proposed ML-based malware detection techniques to secure home routers again DDoS attacks. They used PCA, one-class SVM and an anomaly detector based on n-grams. Home routers in a smart home scenario are the luring targets for malware-based attacks where known vulnerabilities will provide a perfect playground for the attackers. The authors focused on anomaly detection through semi-supervised approaches and the experimental results reported in the paper demonstrate that these classifiers achieve higher detection rate and accuracy.  

Similarly, Esmalifalak et al. \cite{Esmalifalak2017} use SVM and PCA to detect false data injection and stealthy attacks in smart grid. The application of this technique could be easily incorporated into IoT. The authors used two methods, in the first method, SVM is leveraged where labeled data is used for supervised learning for training the SVM whereas in the second methods no training is used. Additionally the authors also used unsupervised learning. These ML techniques are used to isolate the tampered data from the normal data and thus-forth detecting the attacks.

\subsubsection{DL-based malware analysis in IoT}

Pajouh et al. \cite{HADDADPAJOUH2018} proposed an RNN-based DL approach for malware analysis technique in IoT. The authors considered Advanced RISC Machines (ARM)-based applications in IoT. The authors train their models with different existing malware datasets and then test their framework with the new malware.  
Similarly, in another work, Azmoodeh et al. \cite{Azmoodeh2018} aimed at a breed of IoT known as Internet of Battlefield Things (IoBT) and used DL technique to analyze the Operational Code (OpCode) sequence of the devices. The authors leveraged deep eigenspace learning and deep convolutional networks techniques to classify the malware in ARM-compatible IoT applications. In their analysis, the authors used Class-Wise Information Gain technique for features selection where both benignware and malware samples were selected for training. The authors used the OpCode sequences of the selected applications for the classification.  


Karbab et al. \cite{KARBAB2018} proposed MalDozer, a DL-based malware analysis tool for android application framework. The detection framework is based on ANN and tested with both benign and malware applications in the android platform. In essence, MalDozer is based on sequences such as API method calls in android, resource permissions, raw method calls, and so on. Furthermore, the proposed technique also automatically engineers features during the training. On the other hand, Su et al. \cite{Su2018} proposed an image recognition-based DDoS malware detection mechanism in IoT networks. In this solution, the authors first collect and classify two major families of malware, i.e. Mirai and Linux.Gafgyt and then convert the program binaries of the IoT applications to grey-scale images. After that a small-scale CNN is applied to classify the images to goodware and malware.  
Meidan et al. \cite{Meidan2018} used deep autoencoders to detect Botnet attacks in IoT. In this solutions, the authors extract the network behavior and then use deep autoencoders to isolate the anomalous network behavior.  The preceding discussion on machine and deep learning approaches for malware detection in IoT networks show that these approaches both require large amount of data for training and extensive training to detect new unknown malware.   

\subsection{Lessons Learned}

Increasing interest in the interdisciplinary research has led to the increased use of ML analytics and related tools. Various ML techniques such as DL, RL and NN are being used to make IoT devices smart and intelligent. Decrease in the computational costs, improved computing power, availability of unlimited solid-state storage and integration of various technological breakthroughs have made this possible. 

ML is used to create models, that are used to design, test, and train the data sets. These ML algorithms are used to detect possible patterns and similarities in large data sets and can make predictions in new upcoming data. However, the fundamental limitation of ML approaches is that, it mostly needs dataset to learn from, and then the learned model is applied to the real data. This phenomena may not encompass the whole range of features and properties of the data. In this regard, DL techniques have been employed to address the limitations of the ML techniques.

Few consumer related examples, starting from Apple's Siri, Microsoft's Cortana, Amazon's Alexa to Google Photos and from Spotify to Grammarly, are all driven by DL algorithms. Similarly, DL is also used in industrial domains such as financial industry for predicting stock price, health-care industry for re-purposing tested drugs for diseases, and for governmental institutions to get real-time predictions in food predication and energy infrastructure. 

DL, RL, and DRL algorithms are some of the promising research areas leveraged for automated extraction of complex features from large amounts of high dimensional unsupervised data. Despite the fact that recent research in this domain has shown good performance, concrete theoretical and analytic foundations are still missing.  The combined computational capabilities of RL and DL incur computational and storage overhead.  Therefore, despite of their performance, these methods may not be well-suited for the resource-constrained IoT devices and for the moderate size datasets.

Although ML and DL are considered to be efficient techniques for classification and predictions in several tasks, these are not the silver bullets that will cater all the challenges faced by the IoT networks. Furthermore, despite the fact that ML and DL are the latest technological and computational trends, they exhibit distinct limitations and challenges that need to be addressed before its widespread adaptation in IoT networks. Also, like other ML techniques, DL is susceptible to mis- and over-classification. Additionally, RL suffers from instability and divergence when all possible action-reward pairs are not used in a given scenario \cite{Mufti}.

\section{Future Research Challenges} 
\label{sec:challenges}
In this section, we discuss the challenges faced by and future research opportunities in ML and DL techniques for the security of IoT networks.

\subsection{Challenges and Limitations of DL and DRL}

\subsubsection{DL -- one size does not fit all}
DL techniques are very much application-specific where a model trained for solving one problem might not be able to perform well for another problem in the similar domain. The models usually need to be retrained with respective data to be used for other similar problems. This phenomenon might not be a problem for some static networks; however, for the real-time IoT applications, such models will be difficult to use. We believe that more insights are needed for DL techniques to be optimized and used for particular IoT applications.  

\subsubsection{Neural networks are black boxes}
Deep neural networks act like a Blackbox, as we do not know how does any DL model reach a conclusion by manipulating parameters and input data. Just like human brain,  it is impossible to find how the brain works and the obtained solution is the result of embedded neurons at intricately interconnected layers. Similarly in DNN, it is impossible to see how complex is the process of decision-making from one layer is transported to the next. Therefore, it is hard to predict the exact layer for the possible failure and hence becomes unsuitable for those applications in which verification is important.  For instance, if DNNs are used in disease prediction and diagnosis for a patient, solid reasoning for his prediction is required to assure the accuracy of diagnosis. Similarly, attack detection and intrusion detection, for instance, follow the same pattern and the shortcomings of deep networks must be addressed.  

\subsubsection{Longer convergence time}
Bootstrapping is one of the major challenges faced during RL convergence phase. Most of the RL algorithms becomes NP-hard as it takes longer to converge. This convergence problem becomes very dangerous in some situation (e.g. autonomous driving), thus making it unsuitable for real time applications. Furthermore, in case of safety-critical systems, time is essence, and the system cannot accommodate delays. Therefore, more research is needed to design efficient bootstrapping mechanisms. 

\subsubsection{Butterfly effect of ML and DL}
Butterfly effect is a phenomenon where a minute change in the input of a system creates chaos in the output. In this regard, ML and DL are also susceptible to this effect where a slight change in the input data to the learning system will create enormous change in the output which is the learned model. This phenomenon expose ML and DL techniques used in IoT to security attacks where the attackers deliberately change the input data to make the system unstable. As aforementioned, AML uses the noise in the input data where the learned networks behave completely differently. Therefore, it is important to maintain the integrity of the input data which is not an easy task where sheer amount of data with high frequency is created. Another important point is that such attacks are more dangerous since these attacks do not need an access to the system itself. More investigation is needed in this direction to devise integrity mechanism for each domain of the IoT applications. 

\subsubsection{Challenges for DL in the edge}
IoT will leverage the advantages of edge computing and leveraging ML and DL for IoT applications, as aforementioned, will increase the applications and services space. However, the overhead incurred by DL techniques and the sheer amount of data generated by IoT devices, it will be hard to implement DL techniques in the edge devices. Furthermore, the time required for training a deep network also plays an important role where real-time and time-critical applications would not able to take the advantage of DL in the edge. The stability of DL models is also important where newly available information will affect the already trained model. Therefore, more investigation is needed in this direction.

\subsubsection{Vulnerability to security breaches}
Deep learning is expected to empower cybersecurity; however, it may be vulnerable to malicious attacks. As drastic different output can be obtained if input modifications are made to such models. For instance, DL is used to control self-driving vehicles and if malicious user happen to access and change the inputs to the deep learning model, vehicle can be potentially controlled by adversary. Therefore, more in-depth investigation is needed from security standpoint in such systems.


\subsection{Challenges Related to IoT Data}
The heterogeneity of the IoT network enables the production of sheer amount of data with very high frequency from different domains. This huge amount of data from varying data leads to various issues related to data collection, security, dependency and unavailability of appropriate and enough data sets. More in-depth investigation is needed to come up with big data techniques to cater with the volume and heterogeneity of IoT-generated data. 

\subsubsection{Data collection}
Data collection from every domain is not straight forward. For instance, data collected from vehicular sensors is used for the management and maintenance of the vehicles and also for the efficient traffic management. Similarly data generated from smart home appliances and body sensors contain personal information that would easily jeopardize user privacy. Similarly, medical data of the patience collected by the service providers have similar challenges. All the aforementioned data is subject to processing with ML and DL algorithms. However, unbalanced negative and positive data and false positive will have catastrophic consequences on our lives. Therefore, the ML and DL techniques must be matured enough to be used commercially in such sensitive domains. One way would be to develop ML and DL methods that do not solely rely on the previous data and sharply learn from the patterns and able to decide in a way that minimize the side effects of the possible outcomes. This phenomenon is more important in situations when ML and DL are used for security solutions where false positives and true negative will have dire consequences on the networks. One solution could be the context-aware ML and DL solutions. In this regard, more research is needed.

\subsubsection{Proximity effect}
Proximity can play a pivotal role in the data collection. The fact that the Internet is walled-off, adds to the challenges of collecting data in IoT networks. Usually the Internet in such domains has limited access from the outside to mitigate different kinds of attacks; however, it adversely affects the utility of the applications. This phenomenon is also related to the legislation indirectly where the data collection policies must be defined for cross-platform, cross-network, and cross-organization. The storage of these data is also subject to in-depth investigation. For instance, where and how to store the medical records of the patients, whether in public or private cloud, who should have access to such data, how to apply ML and DL algorithms to such data and what level of privacy should be preserved by the ML and DL algorithms. These questions must be further investigated. Furthermore, cloud and fog computing technologies are vulnerable to plethora of attacks that will jeopardize the health of the data and the security of users. 

\subsubsection{Data dependency}
DL algorithms are data hungry and are required to learn progressively from data, and work best if hight quality data is available. Similar to human brain that needs lots of experience to learn first and afterwards is able to make conclusions; it requires huge amount of data to make powerful abstractions. Similarly, DL algorithm can predict about company's stocks after thoroughly learning about historic rise and fall in company's stocks. If enough data is not available, the DL system will fail to make profound predictions. This is due to the fact that DL algorithm works in a systematic way; that is, first learn about the domain and afterwards solve the problem. During the training phase, DL algorithm starts from scratch and needs a huge number of data/parameters to tune/play around. 

\subsubsection{Unavailability of training datasets}
Efficient use of ML and DL solutions need profound datasets that are currently missing.  Furthermore, the rules and policies required for defining the learning strategies still need to be explored. Additionally, authentic datasets from real physical environment are required to analyze and compare the performance of various DL and RL algorithms. To date, efforts have been made to cope with this challenge, but more research is needed in this direction. 



\subsection{Efficiency of Security Solutions}
The degree of sophistication of a security mechanism depends on the capabilities of the device and the system where it is used. The limitations of the IoT devices is a major challenge in applying sophisticated security mechanisms. In the previous section, we discussed the ML- and DL-based security mechanisms; however, the resource constraints create a set-back where a trade-off is needed between the level of security and the capabilities of the IoT devices. Sophisticated security solutions need considerable amount of computing, storage, and communication resources which are the assets in IoT networks. Furthermore, it is also important to determine where to put the logic of ML and DL techniques in the network. Therefore, in-depth investigation is needed for the efficiency of the security mechanisms that use ML and DL techniques. This phenomenon is essential in mission- and time-critical applications. In this regard, low-cost and highly efficient security mechanisms must be investigated for IoT where they can harness the benefits of the ML and DL as well. 

\subsubsection{Complex cyber threats}
IoT networks use resource-constrained devices ranging from home-appliances to personal gadgets. These devices are usually easy targets for cyber attacks. As aforementioned, sheer amount of data is generated by these devices which might be used by ML and DL techniques for different applications. The compromise of these devices will have dire consequences on the outcomes of the applications. It is note that for applications such as smart home, the consequences might not be that critical as compared to critical-infrastructure and medical applications. These applications will not be able to withstand the results from the ML and DL systems as a result of compromised data. It could even endanger human lives. Therefore, it is essential to make sure the device safety and the health of the data that is input to the ML and DL systems. Furthermore, the compromised devices could also be used as bots by the attackers as launching pads for other attacks. Therefore, for ML and DL systems to work in a safe way, it is essential to focus on the security aspects of the IoT devices and on the health of the generated data. In this context, research on the quality of data is a good direction in future.  

\subsubsection{Privacy preservation in ML- and DL-based solutions}
The big data generated by IoT networks is used by analytical systems that leverage ML and DL techniques. These data contain personal and critical information that would not only identify the users but also their behavior and lifestyle. For instance, the data generated by BAN and other health-care related applications might compromise the user privacy and the data from smart home might result in exposing personal lifestyle as well as behavior. Therefore, it is important to make sure that the data used by ML and DL techniques does not put the user privacy at stake. To date, many anonymization techniques have been used that anonymize the data before using it for analytics; however, researches have also shown that the anonymization techniques can be hacked and the training models can be compromised by injecting false data. Learning is the essence of the ML and DL techniques and therefore, feeding the wrong data will fail the purpose of these models. It is, therefore important to investigate data protection and user privacy preservation techniques in ML and DL-based analytics for IoT networks. 

\subsubsection{Adversarial machine learning and IoT}
Machine learning is a double-edge sword where on one hand it nourishes the value of the data, but on the other hand, can be used by the attackers for malicious purposes. Such branch of ML is called Adversarial Machine Learning (AML). In AML, the attackers use the features of the ML to attack the system. For instance, much research has been done by playing with the training parameters and misleading the learning system to learn the opposite of what it is supposed to do. More precisely, perturbation has been used in object recognition where changes into the classifiers causes the system to identify the wrong object. Recently, a one-pixel perturbation was used to fool a DNN \cite{Su2017}. It is therefore important to investigate the role and effects of AML in the IoT networks and address these challenges. 

\subsubsection{Lack of awareness and consumer's confidence in IoT security}
For the successful realization and commercialization of IoT applications, it is essential to increase the consumer stimulation in using IoT services. Consumer satisfaction is of paramount importance for the investors to invest in the IoT. However, recent surveys show that IoT security and privacy are two of the main concerns in using IoT services and the consumers do not seem to feel safe while using IoT services. 
In presence of such concerns, it is extremely important to devise efficient security solutions for IoT applications. More research is needed in this direction that encompass the whole security and privacy spectrum.

\section{Conclusion}
\label{sec:conclusion}

IoT security and privacy are of paramount importance and play a pivotal role in the commercialization of the IoT technology. Traditional security and privacy solutions suffer from a number of issues that are related to the dynamic nature of the IoT networks. ML and more specifically DL and DRL techniques can be used to  enable the IoT devices to adapt to their dynamic environment. These learning techniques can support self-organizing operation and  also optimize the overall system performance by learning and processing statistical information from the environment (e.g. human users and IoT devices). These learning techniques are inherently distributed and do not require centralized communication between device and controller. However, the datasets needed for ML and DL algorithms are still scarce, which makes benchmarking the efficiency of the ML- and DL-based security solutions a difficult task. In this paper, we have considered the role of ML and DL in the IoT from security and privacy perspective. We have discussed  the security and privacy challenges in IoT, attack vectors, and security requirements. We have described different ML and DL techniques and their applications to IoT security.  We have also shed light on the limitations of the traditional ML mechanisms.  Then we have discussed the existing security solutions and outlined the open challenges and future research directions. In order to mitigate some of the shortcomings of machine learning approaches to IoT security, the theoretical foundations of DL and DRL will need to be strengthened so that the performances of the DL and DRL models can be quantified based on certain parameters such as computational complexity, learning efficiency, parameter tuning strategies, and data driven topological self-organization. Furthermore, new hybrid learning strategies and novel data visualization techniques will be required for intuitive and efficient data interpretation.

\bibliographystyle{IEEEtran}
\bibliography{References}
\end{document}